\documentclass[a4paper,11pt]{article}
\pdfoutput=1 

\usepackage{jcappub} 
\usepackage{amssymb}
\usepackage{xcolor}
\usepackage[T1]{fontenc} 
\usepackage{amsmath}
\usepackage{amsfonts}
\usepackage{mathrsfs}
\usepackage[scr=rsfso,cal=zapfc,frak=euler,bb=ams]{mathalfa}
\usepackage{trimclip}
\usepackage{accents}
\newcommand{\halfapprox}{\clipbox{0em 0em 0em 0.225em}{$\approx$}}
\DeclareUnicodeCharacter{2212}{-}

\newcommand{\mystar}[1][black]{\Large\textcolor{#1}{\ensuremath\star}} 

\title{\boldmath Most general isotropic charged fluid solution for Buchdahl model in $\mathscr{F}(Q)$ gravity }

\author[a]{Sourav Chaudhary}
\author[b,1]{Sunil Kumar Maurya
\note{Corresponding author}}
\author[a,2]{Jitendra Kumar\note{Corresponding author}}
\author[c]{Ghulam Mustafa}

\affiliation[a]{Department of Mathematics, Central University of Haryana, Jant-Pali, Mahendergarh, Haryana 123029 India}
\affiliation[b]{Department of Mathematical and Physical Sciences,
College of Arts and Sciences, University of Nizwa, P.O. Box 33, Nizwa 616, Sultanate of Oman}
\affiliation[c]{Department of Physics,
Zhejiang Normal University, Jinhua 321004, People’s Republic of China}

\emailAdd{chaudharysourav192@gmail.com}
\emailAdd{sunil@unizwa.edu.om}
\emailAdd{jitendark@gmail.com}
\emailAdd{gmustafa3828@gmail.com}

\abstract{In this work, we investigated a most general isotropic charged fluid solution for the Buchdahl model via a two-step method in $\mathscr{F}(Q)$-gravity framework for the first time. In this context, a linear function of the form $\mathscr{F}(Q)=\zeta_1 Q+\zeta_2$ and a particular transformation is used to solve the Einstein-Maxwell Equations (EMEs) employing the Buchdahl ansatz: $ e^{\Upsilon(r)}=\frac{\mu(1+\lambda r^2)}{\mu+\lambda r^2}$, where $\zeta_1$, $\zeta_2$, $\lambda$ and $\mu$ are constant parameters. The Schwarzschild de Sitter~(AdS) exterior solution is joined to the interior solution at the boundary to determine the constant parameters. It should be emphasized that, for a given transformation, the Buchdahl ansatz only offers a mathematically feasible solution in the context of electric charge, where pressure and density are maximum at the center and decrease monotonically towards the boundary when $0<\mu<1$. We taken into account the compact star EX01785-248 with $M=(1.3\pm 0.2)M_{\odot}$; Radius $=12.02^{+0.55}_{-0.55}$~km for graphical analysis. 
The physical acceptability of the model in the context of $\mathscr{F}(Q)$ has been evaluated by looking at the necessary physical properties, including energy conditions, causality, hydrostatic equilibrium, pressure-density ratio, etc. Additionally, we predicted the maximum mass limit of different compact objects for various parameter values along with the mass-radius relation. The maximum masses range (1.927 - 2.321)~$M_\odot$ are obtained for our solution. It can be observed that when the coupling parameter $\zeta_1$ for $\mathscr{F}(Q)$  gravity is smaller, then our solution yields massive stars. The present investigation provides novel insights and realistic implications regarding the formation of compact astrophysical objects.}

\begin{document}
\maketitle
\flushbottom

\section{Introduction}
\label{sec:intro}
Modified gravities are significant because they fill some of the gaps in our knowledge of the cosmos and offer an alternate explanation to the General Theory of Relativity (GTR). Using Riemannian geometry, the traditional explanation of general relativity (GR) specifies that the affine connection across the spacetime manifold is metric compatible~\cite{Heisenberg:2018vsk}. However, each manifold can have a variety of affine connections, and these connections can result in distinct but equivalent representations of gravity~\cite{BeltranJimenez:2019esp,Harada:2020ikm}. In contrast to curvature $R$, nonmetricity $Q$ and torsion $T$ are the other two basic geometrical objects, both vanish according to GR's Levi-Civita connection. In addition, one can establish the teleparallel equivalent of GR (TEGR)~\cite{ Maluf:2013gaa} when selecting a relationship that demands nonmetricity and curvature to vanish but relaxes the requirement on torsion. Numerous interesting theories involving modified versions of gravity exist, including teleparallel gravity, unimodular gravity, $\mathscr{F}(R)$, $\mathscr{F}(R,T)$ gravity, $\mathscr{F}(Q)$ gravity, Born-Infeld inspired modifications of gravity, and many more. Furthermore, different contexts can be observed based on these modified metrics, which can be further studied in sequential works~\cite{BeltranJimenez:2017doy,Wang:2021zaz, Errehymy:2022hxt,Kaur:2022gkh,Astashenok:2020qds,Maurya:2020rny,Maurya:2019hds,Nojiri:2017ncd}.\\

One of the more trustable theories for overcoming the limitations of GR is the $\mathscr{F}(R)$ theory. The Lagrangian's scalar curvature can be changed to generate $\mathscr{F}(R)$ gravity. The importance of this idea in combating dark energy has been demonstrated in~\cite{Capozziello:2011et, Sotiriou:2008rp}. The affine connection in GR allows the Ricci scalar $R$ to be nonzero~\cite{BeltranJimenez:2018vdo} while allowing the nonmetricity $Q$ and torsion $T$ to vanish, in a manner analogous to the construction of $\mathscr{F}(T)$ gravity. In an extensive range of theories and geometries, nonmetricity, curvature, and torsion can be combined in many ways to become zero or non-zero. As mentioned previously , we acquire a symmetric teleparallel formulation in GR (STGR) when we choose to assume zero curvature and nonmetricity in the case of $\mathscr{F} (T)$ gravity. Various studies in the subject of STGR have been conducted and are listed as~\cite{Nester:1998mp,Adak:2005cd,BeltranJimenez:2017tkd,BeltranJimenez:2018vdo,Gakis:2019rdd,Li:2022vtn,Li:2021mdp}. To construct $\mathscr{F}(Q)$ gravity, the Lagrangian comprises an arbitrary form of non-metricity scalar. Compared to the fourth-order differential equations in $\mathscr{F}(R)$ theory, the second-order field equations of $\mathscr{F}(Q)$ gravity offer an important advantage. In the present study, the main center of attraction was on $\mathscr{F}(Q)$ gravity. Without taking into account, the affine space-time structure, the modified gravity offers a simple composition of the classical GR, which helps to enhance inertial gravitational interaction. The capability of $\mathscr{F}(Q)$ gravity to interpret the Universe's late-time acceleration without changing scalar fields is its unique characteristic~\cite{Akarsu:2010zm,BeltranJimenez:2019tme,Lazkoz:2019sjl}. The $\mathscr{F}(Q)$ gravity model's viability has been proved in a wide range of domains, including wormhole geometry~\cite{Harko:2018gxr, Xu:2019sbp,DAmbrosio:2020nev, Maurya:2022vsn,Maurya:2023kjx, Banerjee:2021mqk,Kiroriwal:2023nul} and dynamical analysis~\cite{Lu:2019hra}, which covers the cosmological aspects~\cite{Mandal:2020buf}, have been presented in relation to $\mathscr{F}(Q)$ gravity. The mathematical representation of $\mathscr{F}(Q) = Q + Q^n$~\cite{Shekh:2021ule} has been used to investigate holographic dark energy. In this work, the non-vanishing $\mathscr{F}(Q)$ function does not have a Schwarzchild analogous solution when searching for a static spherically symmetric solution.
 Regardless of the extensively accepted, the teleparallel version of GR shares a formulation with both STGR and GR~\cite{Heisenberg:2022mbo}. The primary contrast lies in the fact that in STGR, gravity is allocated to the non-metricity $Q$, whereas curvature and torsion are regarded as zero. Compatible with the non-metricity and zero value of the torsion restriction, a coincident gauge is always applicable~\cite{BeltranJimenez:2017tkd}. This method is unique for the reason that it eliminates the affine connections. Nevertheless, the development of metrics will differ in different coordinate systems if we continue with the incident gauge~\cite{Lu:2019hra,Frusciante:2021sio} and treat the metric as the only fundamental variable in modified STGR theories, such as the $\mathscr{F}(Q)$ theory. 
 
 More importantly, a topic worth discussing is the study of celestial structures in modified theories of gravity. Several research on star structure have been done in the previous few decades. However, Nashed and Capozziello~\cite{Nashed:2021sji} have recently studied the compact object models in $\mathscr{F}(R)$ gravity. They tested the durability of the compact objects using the Tolman-Oppenheimer-Volkoff  (TOV) equation, and their outcome agrees with the observational data of pulsars presenting thermonuclear bursts and millisecond pulsars with a white dwarf companion. Furthermore, the components of metric potential can be rewritten in terms of a compactness parameter $C$ which has to be $C = 2GM/Rc^2$ for physical consistency. The constants that describe the solution are successfully matched by a compactness parameter $C$
which satisfies the constraint $C << 0.5$. In $\mathscr{F}(R)$ gravity, Astashenok and Odintsov~\cite{Astashenok:2020cqq} examined spinning Newton stars with axion fields and talked about stars with various masses and frequencies. On rotating at the same frequency, they found that stars with intermediate masses $M_{\odot} < M < 1.4M_{\odot}$ are more compact. Their findings highlight further noteworthy aspects of stellar configurations in modified gravity with axions, which should be seen in the future. Odintsov and Astashenok~\cite{Astashenok:2020cfv}, investigated realistic neutron stars in axion $R^2$ gravity. They found an increase in the star's mass, independent from central density for a wide range of masses. Additionally,  they predict the possible existence of supermassive compact stars with masses $M\halfapprox$ $2.2-2.3M_{\odot}$ and radii $R\halfapprox$$11km$. In the case of quadratic $f(R)$ gravity, Capozziello et al.~\cite{Capozziello:2015yza} studied the mass-radius relation for static neutron star models. They found greater masses and radii at lower central densities, with an inversion of the behavior around a pivoting $\rho_c$ which mainly depends on the choice of the equation of state. Astashenok et al.~\cite{Astashenok:2020qds, Astashenok:2021peo} demonstrate that the mass-radius relation derived from Extended Theories of Gravity may adequately characterize the neutron star with a mass of 2.67~$M_{\odot}$. They also claimed that under some equations of state, the masses of rotating neutron stars can be greater than 2.67~$M_{\odot}$. In addition, Maurya et al.~\cite{Maurya:2021qye} have used the gravitational decoupling method within the 5D Einstein-Gauss-Bonnet formalism to show the existence of strange star candidates above the traditional maximum mass limit. Further solutions, such as cosmological models, have been studied in $f(T)$, $f(R,T)$ and $f(T,T)$ modified gravity~\cite{Mishra:2020jjk,Salako:2020tno,Ghosh:2020rau,Das:2015gwa}.

A variety of techniques have been researched to create and evaluate physically feasible models of compact objects~\cite{Gupta:2005yps, Kumar:2017zzu}, including the Vaidya-Tikekar ansatz~\cite{Vaidya:1982zz}. Another kind of metric was developed by Buchdahl~\cite{Buchdahl:1959zz}, who put up an important framework for obtaining a physically acceptable ideal fluid solution that is spherically symmetric and has a density that decreases monotonically as it approaches the boundary. The Karmarkar condition was a novel method employed by researchers to solve the field equation for smaller objects. Numerous studies of the stellar structures have been conducted under modified gravity using the Karmarkar condition, See Refs~\cite{Mustafa:2020jln, Mustafa:2020yux, Mandal:2021qhx}. With the isotropic form of matter distribution and the static spherically symmetric metrics, researchers are constantly interested in finding exact solutions to EMF equations. Even though numerous researchers have put forth several answers, and their consequences for observational and theoretical frameworks~\cite{Bhar:2017jbl, Maurya:2017aop, Maurya:2017sjw, Lemos:2014lza, Rahaman:2010mr, Gupta:2012ux, Hansraj:2017bki}. A specific type of gravitational potential with linear EoS that fits quark matter can be specified in~\cite{Komathiraj:2007cw}, in order to achieve the valid Einstein-Maxwell solution. Varela et al.~\cite{Varela:2010mf} have also discovered a novel approach for charged anisotropic materials with linear or nonlinear EoS. The EMF equations were successfully solved by Prasad et al.~\cite{Prasad1} through the appropriate selection of the metric potential, and the Karmarkar condition. Based on the physical analysis of their model, the relativistic stellar structure derived for anisotropic matter distribution is a scientifically plausible model for a compact star with an energy density of around $10^{15} g/cm^3$. By altering the $f (R, T)$-coupling parameter $\chi$, Singh et al.~\cite{Singh1} examined the characteristics of the solution and discovered that $\chi = -1$ produces a stronger EoS than $\chi = 1$. This is because the EoS parameter $\omega$ is larger, the sound velocity is greater, and the $M_{max}$ on the M-R curve is higher for smaller values of $\chi$. The M-R diagram obtained from their solution fits a small number of measured values of compact stars, including PSR J1614-2230, Vela X-1, Cen X-3, and SAX J1808.4-3658.  Consequently, one of the many simplified assertions that have been made to bring together the field equations is the presumption of metric potentials as a result of several reliable data supporting an EoS at maximum densities.

  Motivated by the literature of compact stars in $\mathscr{F}(Q)$ gravity including charged, uncharged, isotropic, and anisotropic models~\cite{Lin,Maurya,Simranjeet,Mustafa1}, we are interested in studying compact stars' configuration of charged isotropic model in $\mathscr{F}(Q)$ gravity. In the present work, we shall investigate the electrically charged isotropic stellar models by using Buchdahl ansatz and a coordinate transformation $e^{\chi(r)}=\mathscr{G}^2(r)$ in the formalism of $\mathscr{F}(Q)$ gravity. It is important to highlight that the Buchdahl ansatz only provides a physically possible solution for the charged condition for a specific transformation, where density and pressure are maximal in the center and monotonically drop towards the boundary when $0 < \mu < 1$. We selected the EXO1785-248 as a charged star candidate to check the physical validity of the stellar model. Using the coordinate transformation, we get the metric function and charge distribution by solving the second-order differential equation with the help of the Gupta-Jasim Two-Step Method. We have demonstrated the physical validity of the charged celestial object, suggesting that the model is well justifiable in the context of $\mathscr{F}(Q)$ gravity.

The paper is presented as follows: In Sec.~\ref{sec:intro}, we provide a brief introduction, Sec.~\ref{sec: field equations} devoted with a summary of $\mathscr{F}(Q)$ gravity with isotropic charged stellar configuration. The system of equations has been resolved with a particular focus on the metric potential, which is Buchdahl~\cite{Buchdahl:1959zz} in the same section. Sec.~\ref{sec: regularity and reality cond.} provides insight into the reality-regular conditions and Sec.~\ref{sec: bound. cond.} includes the boundary conditions that are used to find the constant coefficients. Sec.~\ref{sec: struct. confi.} deals with the structure configuration of compact stellar objects. Subsequently, Sec.~\ref{sec: Modified. Buchdhal.} and Sec.~\ref{sec: TOV.} provides the modified Buchdahl limit of gravitational collapse in the context of $\mathscr{F}(Q)$ gravity and the equilibrium dynamics of the compact stellar object. In Sec.~\ref{sec: MR Curve}, we explore the mass-radius relation of different massive stars. Finally, in Sec.~\ref{sec: Conclusion}, we conclude our findings.

\section{The Action and Field Equations in $\mathscr{F}(Q)$ gravity }
\label{sec: field equations}
The Levi-Civita affine connection is crucial to the standard description and evaluation of GR because, while solving within the space-time manifold, it must be consistent with the metric. It is possible to employ various manifolds, affine connections, and gravity explanations~\cite{Maluf:2013gaa, BeltranJimenez:2019tme}. Torsion ($T$) and nonmetricity ($Q$) are two more important geometric concepts, as demonstrated by the Levi-Civita link, in addition to curvature. With the help of these conditions, a wide range of non-Riemannian geometry-based theories are attainable, including those in which the curvature, torsion, and nonmetricity are non-vanishing. These parameters were used to build the $\mathscr{F}(Q)$ gravity, a Lagrangian with an arbitrary $Q$ form of nonmetricity. An essential component in the universe's expansion is the extension of $\mathscr{F}(Q)$ gravity. The concept of $\mathscr{F}(Q)$ gravity, or symmetric teleparallel gravity, was initially organized by Jimenez et al.~\cite{BeltranJimenez:2019tme}.
In the framework of $\mathscr{F}(Q)$ gravity, a metric space-time is studied, referring to the previously mentioned work of~\cite{Zhao:2021zab}, in which the metric tensors $g_{\alpha\beta}$ and $\Gamma_{\alpha\beta}^\gamma$ are independent entities. The nonmetricity of the connection described above is described by the following equation:
\begin{equation}\label{eq1}
\mathcal{Q}_{\psi\alpha\beta }=\nabla_{\psi}g_{\alpha\beta} =\partial_{\psi}g_{\alpha\beta}-\Gamma^{\gamma}_{\psi\alpha}g_{\gamma\beta}-\Gamma^{\gamma}_{\psi\beta}g_{\alpha \gamma},
\end{equation}
The subsequent three components, in their general form, are obtained by reconstructing this affine relationship into independent parts.
\begin{equation}\label{eq2}
\Gamma_{\alpha\beta}^\gamma=\mathcal{M}_{\alpha\beta}^\gamma+\mathcal{N}_{\alpha\beta}^\gamma+\left\{{ }^\gamma{ }_{\alpha \beta}\right\}
\end{equation}
where  $\left\{{ }^\gamma{ }_{\alpha \beta}\right\}$  is the affine connection, $\mathcal{N}_{\alpha\beta}^\gamma$ denotes disformation \& $\mathcal{M}_{\alpha\beta}^\gamma$ are contortion tensor. The affine connection  $\left\{{ }^\gamma{ }_{\alpha \beta}\right\}$  is solved with metric potential $g_{\alpha\beta}$ as
\begin{equation}\label{eq3}
\left\{{ }^\gamma{ }_{\alpha \beta}\right\} \equiv \frac{1}{2} g^{\gamma \xi}\left(\partial_\alpha g_{\xi \beta}+\partial_\beta g_{\xi \alpha}-\partial_\xi g_{\alpha \beta}\right),
\end{equation}
The contortion $\mathcal{M}^\gamma{ }_{\alpha \beta}$ is defined as, 
\begin{equation}\label{eq4}
\mathcal{M}^\gamma{ }_{\alpha \beta} \equiv \frac{1}{2} \mathcal{Y}^\gamma{ }_{\alpha \beta}+\mathcal{Y}_{(\alpha}{ }^\gamma{ }_{\beta)} .
\end{equation}
The anti-symmetric component of the affine connection is known as the torsion tensor, $\mathcal{Y}^\gamma{ }_{\alpha\beta} \equiv 2 \Gamma_{[\alpha \beta]}^\gamma$ and the disformation $\left(\mathcal{N}_{\alpha \beta}^\gamma\right)$ is defined as:
$$
\mathcal{N}^\gamma{ }_{\alpha \beta} \equiv \frac{1}{2} Q^\gamma{ }_{\alpha \beta}-Q_{(\alpha}{ }^\gamma{ }_{\beta)}
$$

We are presently taking a look at the  conjugate of non-metricity that follows:
\begin{equation}\label{eq5}
\mathcal{H}^\psi{ }_{\alpha \beta}=\frac{-1}{4} Q^\psi{ }_{\alpha\beta}+\frac{1}{2} Q_{(\alpha \beta)}^\psi+\frac{1}{4}\left(Q^\psi-\tilde{Q}^\psi\right) g_{\alpha \beta}-\frac{1}{4} \delta_{(\alpha \beta)}^\psi,
\end{equation}
The following are the independent traces of the aforementioned equations:
$$
Q_\psi \equiv Q_{\psi \alpha}^\alpha \quad \tilde{Q}_\psi \equiv Q^\alpha{ }_{\psi \alpha},
$$
It is now possible to describe the nonmetricity scalar as follows:
\begin{equation}\label{eq6}
Q=-Q_{\psi \alpha \beta} \mathcal{H}^{\psi \alpha \beta} \text {. }
\end{equation}
We employ the electromagnetic field tensor and energy-momentum tensor, which is defined as:
\begin{equation}\label{eq7}
\mathcal{Y}_{\alpha \beta} \equiv \frac{2}{\sqrt{g}} \frac{\left.\delta(\sqrt{-g}) \mathscr{L}_m\right)}{\delta g^{\alpha\beta}}
\end{equation}
\begin{equation}\label{eq8}
\mathcal{E}_{\alpha\beta} \equiv-\frac{2}{\sqrt{-g}} \frac{\delta\left(\sqrt{-g} \mathscr{L}_e\right)}{\delta g^{\alpha\beta}}
\end{equation}
Since the action defined with the use of Lagrange multiples, describes the $\mathscr{F}(Q)$ gravity:
\begin{eqnarray}
&& \hspace{-0.5cm}\mathcal{H}_{\mathcal{G}}=\int \sqrt{-g} d^4 x\Big[\frac{1}{2} \mathscr{F}(Q)+\gamma_\psi^{\xi \alpha \beta} \mathscr{R}_{\xi \alpha \beta}^\psi+\gamma_\psi^{\alpha \beta} \mathcal{Y}_{\alpha \beta}^\psi \nonumber\\&&\hspace{3.5cm}+\mathscr{L}_m+\mathscr{L}_e\Big] .\label{eq9}
\end{eqnarray}
On varying the Eq.~(\ref{eq9}) with respect to the metric $g_{\alpha\beta}$,
we obtain the following field equations
\begin{eqnarray}
&& \hspace{-0.5cm}\mathcal{Y}_{\alpha \beta}+\mathcal{E}_{\alpha\beta}=\frac{2}{\sqrt{-g}} \nabla_\psi\left(\sqrt{-g} \mathscr{F}_Q \mathcal{H}_{\alpha \beta}^\psi\right)+\frac{1}{2} g_{\alpha \beta} \mathscr{F}\nonumber\\&&\hspace{1.5cm}+\mathscr{F}_Q\left(\mathcal{H}_{\alpha \psi, \xi} Q_\beta^{\psi \xi}-2 Q_{\psi, \xi \alpha} \mathcal{H}_\beta^{\psi \xi}\right)\label{eq10}
\end{eqnarray}
The $\mathscr{F}(Q)$'s derivative in regard to $Q$ is represented by the subscript $Q$; that is, $\mathscr{F}_Q \equiv \partial_Q \mathscr{F}(Q)$. 

We define the following equation when we vary the Eq.~(\ref{eq9}) with respect to the affine connection:
\begin{eqnarray}\label{eq11}
\nabla_\rho \gamma_\psi{ }^{\alpha \beta\rho}+\gamma_\psi{ }^{\alpha \beta}=\sqrt{-g \mathscr{F}_Q} \mathcal{Y}_{\alpha \beta}^\psi+\mathcal{D}_\psi{ }^{\alpha \beta}.
\end{eqnarray}
The following equation gives the representation of the hypermomentum tensor density,
\begin{equation}\label{eq12}
\mathcal{D}_\psi{ }^{\alpha \beta}=\frac{-1}{2} \frac{\delta \mathcal{L}_m}{\delta \mathcal{Y}_{\alpha \beta}^\psi .}
\end{equation} 
Now, we reduce the Eq.~(\ref{eq11}), anti-symmetry to the following equation by taking into account the anti-symmetry  of $\alpha$ and $\beta$ in the Lagrangian multiplier coefficients:
\begin{equation}\label{eq13}
\nabla_\alpha \nabla_\beta\left(\sqrt{-g} \mathscr{F}_Q \mathcal{H}_\psi^{\alpha\beta}+\mathcal{D}_\psi{ }^{\alpha \beta}\right)=0,
\end{equation}
if we consider $\nabla_\alpha \nabla_\beta \mathcal{D}_\psi{ }^{\alpha \beta}=0$ we get:
\begin{equation}\label{eq14}
\nabla_\alpha \nabla_\beta\left(\sqrt{-g} \mathscr{F}_Q \mathcal{H}_\psi^{\alpha \beta}\right)=0 .
\end{equation}
Without torsion and curvature, the affine connection has the following form:
\begin{equation}\label{eq15}
\Gamma_{\alpha \beta}^\psi=\left(\frac{\partial x^a}{\partial \varpi^\gamma}\right) \partial_\alpha \partial \beta \varpi^\gamma .
\end{equation}
If we take into account a specific coordinate selection, named a coincident gauge
 such that $\Gamma_{\alpha \beta}^\psi=0$. the non-metricity takes the form:
\begin{equation}\label{eq16}
Q_{\psi \alpha \beta}=\partial_\psi g_{\alpha \beta}
\end{equation}
With the use of this unique coordinate, the computation becomes simpler because the metric is the only remaining basic variable. Even though, the action is not invariant w.r.t diffeomorphism, with the exception of STGR~\cite{Zhao:2021zab}. This problem is solved by using the covariant formulation of the $\mathscr{F}(Q)$ gravity. Since the relationship in Eq.~(\ref{eq15}) is inertial, we may find the affine connection in the non-appearance of gravity by applying the covariant formulation~\cite{BeltranJimenez:2019tme}. However, $\mathscr{F}(Q)$ gravity would be severely constrained by the off-diagonal constituent of the coincident gauge's field equations, contributing us with nontrivial forms of $\mathscr{F}(Q)$ gravity formalism.

The formulation of static and spherically symmetric geometry in polar coordinates is shown below. A generic static space-time with spherical symmetry takes the following metric:
\begin{equation}\label{eq17}
ds^{2}=-e^{\chi(r)}dt^{2}+e^{\Upsilon(r)}dr^{2}+r^{2}d\sigma^{2},
\end{equation}
where $d\sigma^{2}=d\theta^{2}+sin^{2}\theta d\phi^{2}$ and $\chi(r)$ \& $\Upsilon(r)$ are unknown functions of radial coordinate $r$.\\
The nonmetricity scalar $Q$ is expressed in terms of $r$, and we can compute the following equation by inserting equation Eq.~(\ref{eq17}) into equation Eq.~(\ref{eq10}),
\begin{eqnarray}\label{eq18}
{Q}=\frac{-2~\chi^\prime(r) }{r~e^{\Upsilon(r)}}-\frac{e^{-\Upsilon(r)}}{r^2},
\end{eqnarray}
where $\prime $ denotes the derivative w.r.t $r$.
The energy-momentum tensor for an isotropic fluid with spherically symmetric geometry is given by,
\begin{equation}\label{eq19}
\mathcal{Y}_{\alpha \beta}=\left(\mathcal{P}+\rho\right) u_\alpha u_\beta+\mathcal{P} g_{\alpha \beta}.
\end{equation}
Here, $u_\alpha$ represents a four-velocity vector. $\rho(r)$ denotes the energy density, $\mathcal{P}(r)$ is the pressure which is measured in regard to co-moving fluid four-velocity $u^\psi=e^{-\frac{\beta}{2}}\delta_{0}^\psi$.
Similarly, the electromagnetic stress-energy tensor $\mathcal{E}_{\alpha\beta}$ is described as,
\begin{equation}\label{eq20}
  \mathcal{E}_{\alpha\beta}=\big(2\mathcal{F}_{\alpha p}\mathcal{F}_{\beta p}-\frac{1}{2}g_{\alpha\beta}\mathcal{F}_{pq}\mathcal{F}^{pq} \big), 
\end{equation}
Since $\mathcal{F}_{\alpha\beta}=\mathcal{A}_{\alpha\beta}-\mathcal{A}_{\beta\alpha}$ expresses the electromagnetic filed tensor and read as $\mathcal{F}_{\alpha\beta,p}+\mathcal{F}_{p\alpha,\beta}+\mathcal{F}_{\beta p, \alpha}=0$ \&  $(\sqrt{-g}\mathcal{F}^{\alpha\beta})_{,\beta}=\frac{1}{2}\sqrt{-g}j^\alpha$. Here $A_\alpha$ is the four potentials of electromagnetic field and $j$ is the 4-current densities.\\

Only a few non-vanishing components satisfied the assumption that there is a static, spherically symmetric electric field
 $\mathcal{F}_{01}=-\mathcal{F}_{10}$. One can get an electric field $\mathcal{E}$ from Eq.~(\ref{eq20}) as,
\begin{equation}\label{eq21}
\mathcal{E}(r)=\frac{e^{\Upsilon+\chi}}{2r^2}\int_0^r e^{\Upsilon(r)}\sigma(r)r^2 dr=\frac{q}{r^2},
\end{equation}
where $\sigma(r)$ is charge density and $q(r)$ is the total electric charge.
With the help of Eq.~(\ref{eq10}) and Eq.~(\ref{eq19}), the equation of motion for an isotropic fluid is given as, 
\begin{eqnarray}
&& \hspace{-0.5cm}\rho(r)+\mathcal{E}^2(r)=-\frac{\mathscr{F}}{2}+\mathscr{F}\Bigg[\frac{1}{r^2}+Q+\frac{1}{r~e^{\Upsilon}}\left(\chi^{\prime}+\Upsilon^{\prime}\right)\Bigg], \\ \label{eq22}
&& \hspace{-0.5cm}\mathcal{E}^2(r)+\mathcal{P}(r)=\frac{\mathscr{F}}{2}-\mathscr{F}\Bigg[e^{-\Upsilon}\Bigg(-\frac{\chi^{\prime \prime}}{2}-\left(\frac{\chi^{\prime}}{4}+\frac{1}{2 r}\right)\left(-\chi^{\prime}+\Upsilon^{\prime}\right)\Bigg)+\frac{Q}{2}\Bigg], \\ \label{eq23}
&& \hspace{-0.5cm}-\mathcal{E}^2(r)+\mathcal{P}(r)=\frac{\mathscr{F}}{2}-\mathscr{F}\Bigg[Q+\frac{1}{r^2}\Bigg], \\ \label{eq24}
&& \hspace{-0.5cm}0=\frac{\cot \theta}{2} Q^{\prime} \mathscr{F}_{Q Q} \label{eq25} .
\end{eqnarray}
Zhao~\cite{Zhao:2021zab} has studied the space-time which is static and spherically symmetric and have a coincident gauge. If we presume that the affine connection in the specified coordinate system is zero, then the off-diagonal component of Eq.~(\ref{eq9}) can be expressed as follows: the extra prerequisite of the $\mathscr{F}(Q )$ gravity in which the vacuum solutions must satisfied, i.e. $\mathcal{Y}_{\alpha \beta}=0$,  \\
\begin{equation}\label{eq26}
\frac{\cot \theta}{2} Q^{\prime} \mathscr{F}_{Q Q}  =0.
\end{equation}
\begin{figure*}[!htp]
    \centering
\includegraphics[height=6cm,width=7cm]{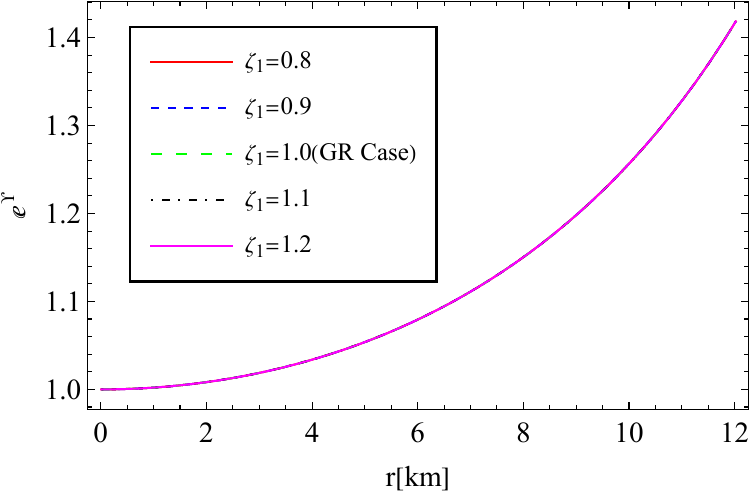} ~~~~~~~~~\includegraphics[height=5.5cm,width=8cm]{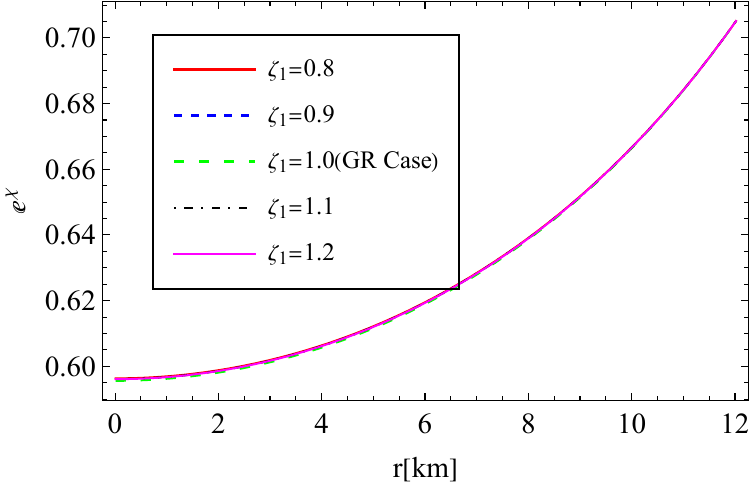}`
    \caption{Display of metric potential - $e^{\chi}$~(left) and  metric potential - $e^{\Upsilon}$~(right) with $\mu=0.0000294  $, $\lambda=-0.00000006$, $\varpi=7.368853$ for different values of $\zeta_1$ }
    \label{fig1}
\end{figure*}
When the equations of motion and Eq.~(\ref{eq10}) are combined in this problem, the outcome is  $\mathscr{F}_{QQ}=0$. The function  $\mathscr{F}(Q )$ should be linear as a result. Inconsistent equations of motion and solutions will result from selecting nonlinear values for the function  $\mathscr{F}(Q )$, especially when  $\mathscr{F}(Q=Q^2 )$ has to be considered. It is evident that an equation of motion including a nonlinear function of the $\mathscr{F}(Q )$ gravity will not yield a consistent value. A more generalized form of the spherically symmetric metric is required in the situation of a coincident gauge if we want to examine and interpret the nonlinear form of $\mathscr{F}(Q )$ gravity. Preferably refer to \cite{Zhao:2021zab} for a detailed analysis of this problem. In current investigation, the spherically symmetric coordinate system Eq.~(\ref{eq17}) which agrees with the affine connection $\mathcal{Y}_{\epsilon \beta}^{\alpha}=0$. Here, in the present problem, we have studied a linear function with $\mathscr{F}_{QQ}=0$. Thus, we employ the function $\mathscr{F}(Q)$, which is expressed in the following manner, to compute the field equation,
\begin{equation}\label{27}
 \mathscr{F}_{Q Q}=0 \Longrightarrow \mathscr{F}(Q)=\zeta_1 Q+\zeta_2,
\end{equation}
where $\zeta_1$ and $\zeta_2$ are constant of integration.
So, with the help of Eq.~(\ref{27}), we get the required filed equations of $\mathscr{F}(Q)$ gravity,
\begin{eqnarray}
\label{eq28}
&& \hspace{-0.5cm}\rho(r)+\mathcal{E}^2(r)=\frac{1}{2 r^2}\Big[2 \zeta_1+2 e^{-\Upsilon} \zeta_1\left(r \Upsilon^{\prime}-1\right)-r^2 \zeta_2\Big], \\ 
 \label{eq29}
&& \hspace{-0.5cm}\mathcal{P}(r)-\mathcal{E}^2(r)=\frac{1}{2 r^2}\Big[-2 \zeta_1+2 e^{-\Upsilon} \zeta_1\left(r \chi^{\prime}+1\right)+r^2 \zeta_2\Big], \\
 \label{eq30}
&& \hspace{-0.5cm}\mathcal{P}(r)+\mathcal{E}^2(r)=\frac{e^{-\Upsilon}}{4 r}\Big[2 e^\Upsilon r \zeta_2+\zeta_1\left(2+r \chi^{\prime}\right)\left(\chi^{\prime}-\Upsilon^{\prime}\right)+2 r \zeta_1 \chi^{\prime \prime}\Big] .
\end{eqnarray}
Our current goal is to solve Eqs.~(\ref{eq28}-\ref{eq30}) in the prior mentioned system. There are five unknowns in all for the system of equations: $\rho(r)$, $\mathcal{P}(r)$, $q(r)$, $\chi(r)$, and $\Upsilon(r)$. It is now crucial to introduce a metric potential that specifies one of the unknown functions or to assuming an EoS that establishes a relationship between pressure and density and provides analytical solutions.\\
The investigation provided in this work takes into account a reasonable selection for Buchdahl's \cite{Buchdahl:1959zz} ansatz. We select the value for the metric function $e^\Upsilon$,\\
\begin{equation}\label{eq31}
    e^{\Upsilon(r)}=\frac{\mu(1+\lambda r^2)}{\mu+\lambda r^2}, 
\end{equation}
where $\lambda<0$ and $\mu$ is arbitrary constant. Even with its extreme simplicity, the model guarantees the reality and regularity conditions at the sphere's center while meeting the physical requisites of a real star. We obtain the metric function taken into consideration by Tikekar and Vaidya~\cite{Vaidya:1982zz} when $\lambda=-\mu/R^2$. Thus the model provides a fascinating geometric interpretation that deviates from the 3-space geometry's uniqueness.\\

Now, defining a new coordinate transformation $e^{\chi(r)}=\mathscr{G}^2(r)$, the field Eqs.~(\ref{eq28})-(\ref{eq30})
take the following form,

\begin{eqnarray}\label{eq32}
&& \hspace{-0.5cm}\frac{\mathscr{G}^{\prime \prime}}{\mathscr{G}}-\frac{\mathscr{G}^{\prime}}{r \mathscr{G}}+\frac{\lambda(\mu-1) r\left(\lambda r-\frac{\mathscr{G}^{\prime}}{\mathscr{G}} \right)}{\left(\mu+\lambda r^2\right)\left(1+\lambda r^2\right)}=\frac{2 q^2\big[\mu\left(1+\lambda r^2\right)\big]}{r^4(\mu+\lambda r^2)}, ~~~~~
\end{eqnarray}
In addition, it is now straightforward to implement the transformation 
 (Gupta-Jasim Two-step method~\cite{Gupta20}):
\begin{equation}\label{eq33}
\eta(r)=\sqrt{\frac{\lambda r^2+\mu}{-\mu+1}}, \quad \text { and } \quad \mathscr{G}=\left(\eta^2+1\right)^{1 / 4} \mathcal{W} .
\end{equation}
With the help of Eq.~(\ref{eq32}) and using the transformation Eq.~(\ref{eq33}) allows us to rewise the $2^{nd}$ order differential equation in simplest form, which is
\begin{equation}\label{eq34}
\frac{d^2 \mathcal{W}}{d \eta^2}+\mathscr{B} \mathcal{W}=0,
\end{equation}
where for notation ease, we use
\begin{equation}\label{eq35}
\mathscr{B}=-\frac{1}{\eta ^2+1}\Bigg[-\frac{5}{4 \left(\eta ^2+1\right)}+\mu  \left(\frac{2 q^2 \left(\lambda  r^2+1\right)}{\lambda ^2 r^6}-1\right)+\frac{7}{4}\Bigg].   
\end{equation}
In order to gain additional understanding and quickly tackle to  Eq.~(\ref{eq34}), we set $\mathscr{B}$ as,
\begin{equation}\label{eq36}
\mathscr{B}=-\frac{2}{\eta\left(\eta +\frac{1}{\varpi^2}\right)},
\end{equation}
where $\varpi$ is a positive constant. Now the relation Eq.~(\ref{eq35}) and Eq.~(\ref{eq36}) conduct to characterize   the total charge given as,
\begin{eqnarray}\label{eq37}
&& \hspace{-0.5cm}\frac{q^2}{r^4}=\frac{1}{2 \mu  \left(\lambda  r^2+1\right)^2}\Bigg[\zeta _1 \lambda ^2 r^2 \Bigg(\frac{2 \varpi ^2 \left(\eta ^2+1\right)}{\varpi ^2 \eta ^2+\eta }+\frac{5}{4 \left(\eta ^2+1\right)}+\mu -\frac{7}{4}\Bigg)\Bigg] .
\end{eqnarray}
Substituting Eq.~(\ref{eq36}) with Eq.~(\ref{eq34}) yields a fresh version of equation, given by
\begin{equation}\label{eq38}
\left(\eta^2\varpi^2+\eta\right) \frac{d^2 \mathcal{W}}{d \eta^2}-2 \varpi^2 \mathcal{W}=0 .
\end{equation}
Upon solving the differential equation mentioned above, we obtain the value of $\mathcal{W}$ as,
\begin{eqnarray}\label{eq39}
&& \hspace{-0.5cm}\mathcal{W}(\eta)=A \Big[-2 \varpi ^4 \eta^2 \log (\eta)+2 \varpi ^4 \eta^2 \log \left(\varpi ^2 \eta+1\right)-2 \varpi ^2 \eta-2 \varpi ^2 \eta \log (\eta)\nonumber\\&&\hspace{4.5cm}+2 \varpi ^2 \eta \log \left(\varpi ^2 \eta+1\right)-1\Big]+B~ \eta \left(\varpi ^2 \eta+1\right),
\end{eqnarray}
where A and B are the integration constants. On using Eq.~(\ref{eq39}) in Eq.~(\ref{eq33}), we get the solution for $\mathscr{G}$.
\begin{eqnarray}\label{eq40}
&& \hspace{-0.5cm}\mathscr{G}(\eta)=\left(1+\eta^2\right)^{1 / 4}\Big[A \Big(-2 \varpi ^4 \eta^2 \log (\eta)+2 \varpi ^4 \eta^2 \log \left(\varpi ^2 \eta+1\right)-2 \varpi ^2 \eta-2 \varpi ^2 \eta \log (\eta)\nonumber\\&&\hspace{4.5cm}+2 \varpi ^2 \eta   \log \left(\varpi ^2 \eta+1\right)-1\Big)+B~ \eta \left(\varpi ^2 \eta+1\right)\Big].
\end{eqnarray}
\begin{figure*}
    \centering
\includegraphics[height=6cm,width=7cm]{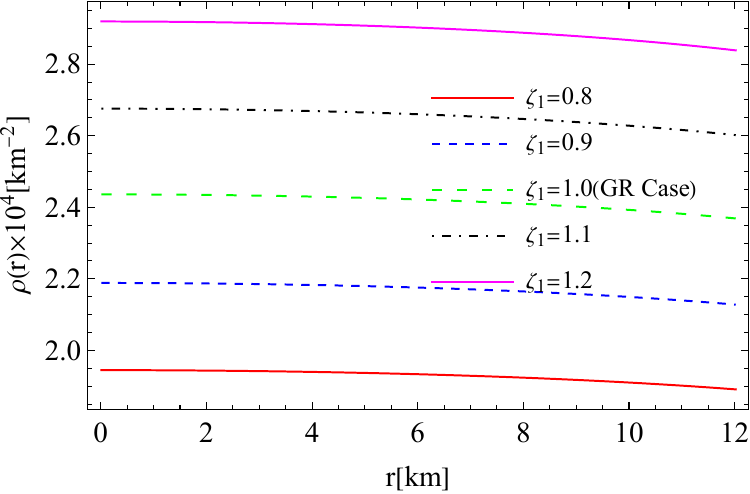}~~~~~~~~~\includegraphics[height=6cm,width=7cm]{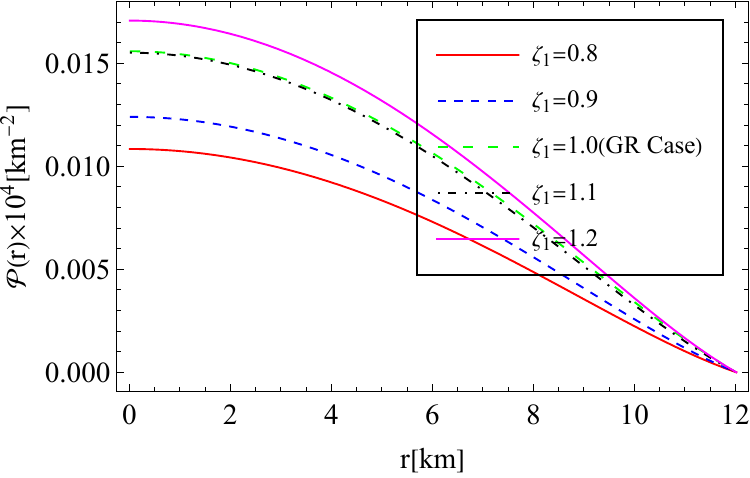}`
    \caption{Display of energy density - $\rho$~(left panel) and  pressure - $\mathcal{P}(r)$~(right panel) with $\mu=0.0000294  $, $\lambda=-0.00000006 $, $\varpi=7.368853$ for different values of $\zeta_1$ }
    \label{fig2}
\end{figure*}
Ultimately, the complete solution of Einstein-Maxwell system  Eqs.~(\ref{eq28})-(\ref{eq30})
\begin{eqnarray}\label{eq41}
&& \hspace{-0.5cm}   \rho(r)= -\frac{\zeta_1 \lambda^2 r^2}{2 \mu \left(\lambda r^2+1\right)^2} \Bigg[\frac{2 \varpi ^2 \left(\lambda r^2+1\right)}{(1-\mu) \left(\eta-\varpi ^2 \eta^2 \right)}+\frac{-3 \lambda r^2-5 \mu+2}{4 \lambda r^2+4}+\mu-1\Bigg]\nonumber\\&&\hspace{5cm}+\frac{2 \zeta_1 \lambda (\mu-1)}{\mu \left(\lambda r^2+1\right)^2}+\frac{\zeta_1 \lambda (\mu-1)}{\lambda \mu r^2+K}-\frac{\zeta_2}{2},
\end{eqnarray}
\begin{eqnarray}\label{eq42}
 && \hspace{-0.5cm}  \mathcal{P}(r)=\frac{\lambda(\mu+\lambda r^2)^2\zeta_1(\eta_{1}+\eta_{2}-\eta_{3}+\eta_{4})}{(\mu-1)^2\mu(1+\lambda r^2)^2\eta^{3/2}(\eta_{5}+\eta_{6}+\eta_{7})}+\frac{2 \zeta_1 \lambda (\mu-1)}{K \left(\lambda r^2+1\right)^2}-\rho.  
\end{eqnarray}

where $\eta_1=-8 \varpi ^2 A-A \eta+A \mu \eta-10 \varpi ^2 A \lambda r^2-2 \varpi ^2 A \mu+3 B \lambda r^2+B \mu+2 B$\\

$\eta_2=5 \varpi ^2 B \lambda r^2 \eta+4 \varpi ^2 B \eta+\varpi ^2 B \mu \eta $.\\

$\eta_3= \varpi ^2 A \left(\lambda r^2 \left(5 \varpi ^2 \eta+3\right)+\varpi ^2 \mu \eta+4 \varpi ^2 \eta+\mu+2\right) $.\\

$\eta_4=\left.2 \varpi ^2 A\left(\lambda r^2 \left(5 \varpi ^2 \eta+3\right)+\varpi ^2 \mu \eta+4 \varpi ^2 \eta+\mu+2\right)\right)$.\\

$\eta_5=B \eta \left(\varpi ^2 \eta+1\right) $.\\

$\eta_6=A \left(\varpi ^4 \eta\log \left(\eta\right)-2 \varpi^2 \eta-2 \varpi ^2 \eta \log \left(\eta\right)-1\right)$.\\

$\eta_7=A \left(2 \varpi ^2 \eta \log \left(\varpi ^2 \eta+1\right)-2 \varpi ^4\eta\log \left(\varpi ^2 \eta+1\right)\right)$.\\

$\eta=\sqrt{\frac{\lambda r^2+\mu}{-\mu+1}}$.\\

We illustrate the outcome in form of different values of the arbitrary constants $\mu$ \& $\lambda$ as we examine the structure of charged spheres composed of perfect fluids. Thus the function of the star's radius, we plot all surface parameters of the stars: density, pressure, density-pressure ratio, and total charge $Q$. Thus, Fig.~\ref{fig2} illustrates a crucial point: in the interior solution, pressure and density are greatest in the center and monotonically decrease towards the boundary. It is the steady stellar structure Buchdahl criteria. Even when this argument is presented in terms of anisotropy and isotropy, one may still see that pressure and density are rising towards the boundary if $\lambda < 0$ and decreasing if $\lambda > 0$ for $0 < \mu < 1$. However, within a certain range of the matter coupling parameters, the density profile is likewise monotonically increasing in the case of the modified theory of gravity, specifically $f (R, T)$ gravity. It indicates that more stable stellar structures, where Coulomb repulsion opposes gravitational attraction, result from charged solutions of the Einstein-Maxwell field equations. The main problem is that, under the current methodology, a feasible sound solution is provided for charged isotropic context, where $0 < \mu < 1$ \& $\lambda < 0$.

   Furthermore, we confirm that pressure diminishes at the boundary in relation to the radial coordinate r (refer to Fig.~\ref{fig2}). It is shown that the electric field connected to  Eq.~(\ref{eq37}), disappears at $r=0$. For various compact stars in Fig.~\ref{fig3}, we display the radius of the generated spheres as a function of the charge distribution. This indicates that the disappearance of the electric field remains positive and regular throughout the sphere at the center of a spherically symmetric charge distribution.

\section{Regularity and reality conditions for the well-behaved model in $\mathscr{F}(Q)$  gravity}
\label{sec: regularity and reality cond.}

The regularity requirements for the $\mathscr{F}(Q)$ gravity model, which must be met for the model to behave accurately are as follows:\\

(i) Finite and positive behavior is required for the density and pressure at the center of the star. This ensures that the compact star does not undergo geometric collapse and that the solution is not affected by singularities. Additionally, the metric potential $e^{\Upsilon}$ and $e^{\chi}$ are positive and non zero.

~~~~~~~~~~~~~~~~~~~~~~~~~$e^{\chi}>0$ and $\big[e^{\Upsilon}\big]_{r=0}=1$, $\rho_{0}>0$ and $\mathcal{P}_{0}>0$.

(ii) The pressure to density ratio, or $\frac{\mathcal{P}}{\rho}$, of the star needs to be positive and have a value lower than 1.
$$0<\frac{\mathcal{P}}{\rho}<1$$.

(iii) At the center of the star, the causality requirement must be satisfied.
$$0<\Bigg(\frac{d\mathcal{P}}{d\rho}\Bigg)_{r=0}\leq 1$$.

(iv) With a rise in the value of $r$,  the factor $\frac{d\mathcal{P}}{d\rho}$ must be positive and increase monotonically.

(v) The solution exhibits positive and monotonically decreasing values of $\frac{\mathcal{P}}{\rho}$ as the values of r increase.

(vi)As $r$ increases, the pressure and density values decrease monotonically and remain positive.

\begin{table*}
\centering
\caption{The numeric data of the constants $A$, $\mu$ and $\lambda$ for different values of $\zeta_1$ \&  $\zeta_2$.}
\begin{tabular}{c|c|c|c|c}
\hline$\zeta_1$ & $\zeta_2$ & $A$ & $\mu$ & $\lambda$ \\
\hline 0.8 & 0.00002 & -0.00231618 & 0.0000294 & -0.00000006 \\
0.9 & 0.00002 & -0.00231624 & 0.0000294 & -0.00000006 \\
1.0 & 0.00000 & -0.00231692 & 0.0000294 & -0.00000006 \\
1.1 & 0.00002 & -0.00231626 & 0.0000294 & -0.00000006 \\
1.2 & 0.00002 & -0.00231629 & 0.0000294 & -0.00000006 \\ 
\hline
\end{tabular}
\label{tab1}
\end{table*}

\section{Boundary Conditions}
\label{sec: bound. cond.}
We examine an appropriate boundary condition in this section for solving the EMF equations in $\mathscr{F}(Q)$ gravity. On matching the interior and external spacetimes, we achieve the solution. Schwarzchild AdS solution can be used to solve the viable exterior solution in $\mathscr{F}(Q)$ gravity, as demonstrated by numerous evidence found in recent studies such as \cite{Wang:2021zaz}. The provided solution is as follows:
\begin{eqnarray}\label{eq43}
&& \hspace{-0.7cm}    d s_{+}^2=  -\Bigg[-\frac{2 \mathcal{M}}{R}+1-\frac{\wedge}{3} R^2+\frac{Q^2}{R^2}\Bigg] d t^2 +{\Bigg[1-\frac{2 \mathcal{M}}{R}-\frac{\wedge}{3} R^2 +\frac{Q^2}{R^2}\Bigg]}^{-1/2} d r^2\nonumber\\&&\hspace{6.5cm}+r^2\left(\sin ^2 \theta d \phi^2+d \theta^2\right),
\end{eqnarray}
where $Q$ is the total charge of the fluid sphere, $\mathcal{M}$ is the total gravitational mass and $\Lambda$ represents cosmological constant. The aforementioned equation allows us to match the interior metric to search out the values of the many constants that are included in the problem. Using the parameters that Darmois-Israel derived in their research \cite{Vaidya:1982zz,Israel:1966rt}, the Schwarzchild AdS metric has been evaluated at surface when $r = R$.\\
\begin{equation}\label{eq44}
\begin{gathered}
\Bigg[1-\frac{2 \mathcal{M}}{R}-\frac{\wedge}{3} R^2+\frac{Q^2}{R^2}\Bigg]=e^{\chi(R)}, \\
\Bigg[1-\frac{2 \mathcal{M}}{R}-\frac{\wedge}{3} r^2+\frac{Q^2}{R^2}\Bigg]=e^{-\Upsilon(R)}, \\
\mathcal{P}(R)=0. 
\end{gathered}
\end{equation}
It is noteworthy that the cosmological constant $\wedge$ is dependent upon the constants $\zeta_1$ and $\zeta_2$, which are expressed as follows: $\wedge=\zeta_2 / 2 \zeta_1$. The present stellar model is not much affected by the value of $\wedge$ since the cosmological constant, which is prescribed to have a significant impact on the cosmological problem, is insignificant in the present situation. The cosmological constant $\wedge$, which is roughly $10^{-46} / \mathrm{km}^2$, is determined by existing numerical evidence to be such a minute number. In the current study, this value is approximately zero. For this work, $\zeta_2$ has been set to 0.00002 throughout the analysis of the stellar configuration. \\

The value of constant $A$ determined in this work is given by
\begin{eqnarray}\label{eq45}
  A=-\frac{B(\mu+\lambda R^2)~(\eta_{26}+\eta_{27}) } {\eta_{28}}.  
\end{eqnarray}
\section{Structure Configurations of compact stellar objects}
\label{sec: struct. confi.}
A star must match up to the essential physical and mathematical requirements of the compact star to be considered relevant. In this instance, we examine the outcomes that arise from transformation and metric potential values to characterize the compact star's interior and exterior components. The various parameters that were used to test the compact star are covered in the section. The feasibility of a charged isotropic solution taking into account $\mathscr{F}(Q)$ gravity is examined in the section that follows. For various values of $\zeta_1 =$ 0.8, 0.9, 1.0, 1.1, 1.2, we test the solution. 
\vspace{1cm}
\subsection{The dynamics of the density,  pressure, and charge of compact stars}
The pressure and density of a compact star in the $\mathscr{F}(Q)$  gravity must decrease monotonically to meet the regularity and reality requirements. The pressure of the star eventually reaches zero at $r = R$, as demonstrated by the star Fig.~\ref{fig2}. It is also remarkable that the energy density and the pressure have mathematical values significantly influenced by parameter $\zeta_1$. As the values of $\zeta_1$ rise, we observe a consistent rise in the energy density and pressure values. The star's core is more compact than its outer layer, as seen by the behavior where the density is highest in the center and decreases towards it. An essential prerequisite for a stable, realistic stellar object showing up of a hydrostatic force feature that neutralizes the gravitational pull to avoid gravitational collapse.\\
\begin{figure}[!htp]
    \centering
\includegraphics[height=5cm,width=7cm]{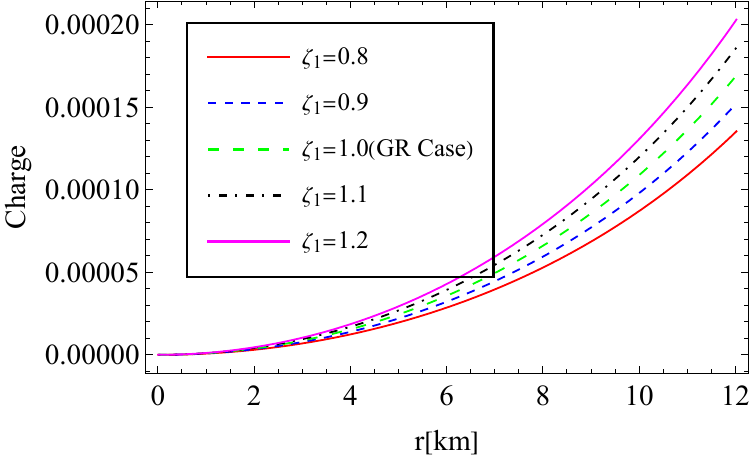}`
    \caption{Profile of electric charge versus `r'}
    \label{fig3}
\end{figure}
Fig.~\ref{fig3} depicts that $0<\mathcal{E}^2 $ across the matter and $0=\mathcal{E}^2_{|r=o} $ in the center, demonstrating that our system is electrically charged. Additionally, after reviewing all the data that is currently available, we will try to justify the amount of charge. According to the literature~\cite{Varela:2010mf}, stars with dissipating net charges are made up of fluid elements at the fluid-vacuum interface with limitless, appropriate net charge densities. The total net charge of these elements can reach enormous values like $10^{19}$ C. Furthermore, the literature~\cite{Ray:2003gt} discusses the consequence of charge in compact objects by considering the maximum amount of charge that they can contain. This suggests, through numerical analysis, that the global balance of forces may permit the presence of a larger amount of charge, such as $10^{20}$ C in a neutron star. In the nearby vicinity of the star, the charge graph must exhibit an increasing nature with positive behavior.
\subsection{Energy Conditions}
Since ideal fluid distributions of matter sustain stellar formations, we will define the energy conditions by classical field theories of gravitation. It is reasonable to believe that there is a relationship that requires pressure and matter density to conform to specific constraints. When determining the significance of charged matter, whether it is feasible or not, energy constraints~\cite{hawking} are used as a technique. The expression form for the Strong, Weak, Null, and Dominant energy conditions (SEC, WEC, NEC, and DEC) for charged objects is as follows.
\begin{itemize}
    \item \textbf{Null Energy Condition}:
    
    $0\leq\mathcal{E}^{2}+\rho$.
    \item \textbf{Weak Energy Condition}: 
    
    $0\leq\rho+\mathcal{P}$,~~$0\leq\rho+2\mathcal{E}^2+\mathcal{P}$.
    \item \textbf{Strong Energy Condition}: 
    
    $0\leq\rho+3\mathcal{P}+\mathcal{E}^2$.
\end{itemize}

\begin{figure*}[!htp]
    \centering
    \includegraphics[height=5.5cm,width=6.5cm]{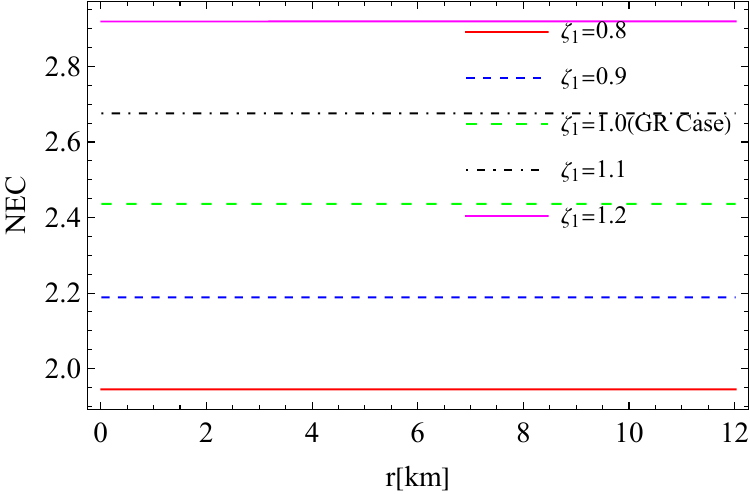}~~~~~~\includegraphics[height=5.5cm,width=6.5cm]{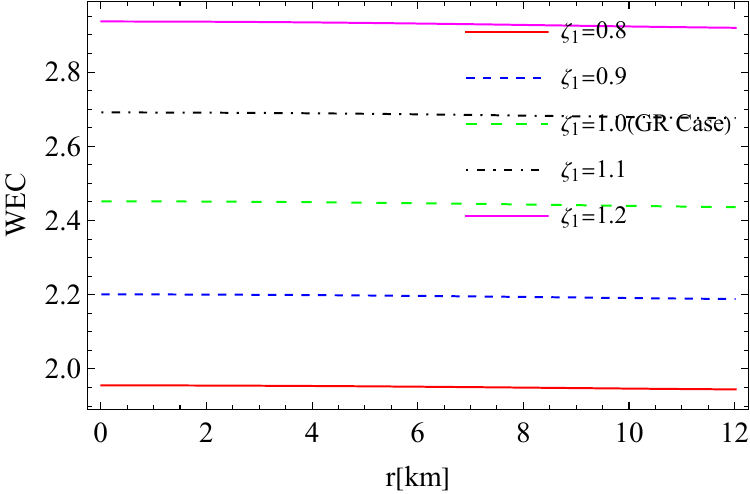}
    
    \includegraphics[height=5.5cm,width=6.5cm]{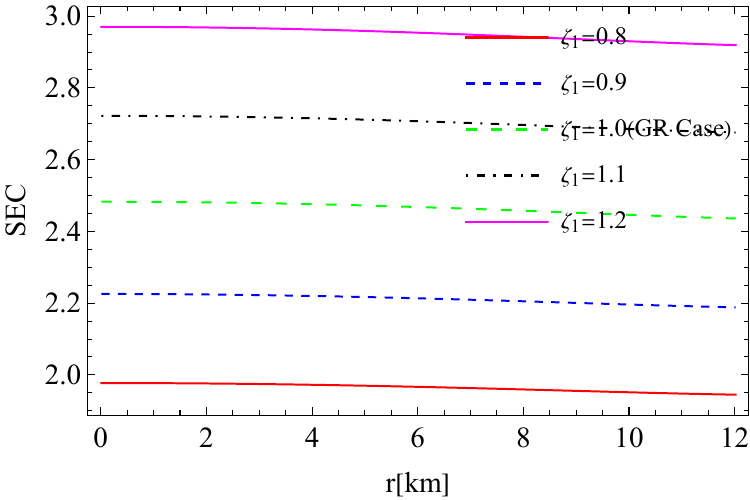}
    \caption{Behavior of energy conditions against `r'.}
    \label{fig4}
\end{figure*}
The electrical charge and realistic charged matter composition of our system of star objects are guaranteed by the positive conduct of energy conditions under investigation See Fig.~\ref{fig4}. 


\subsection{Causality condition}
\begin{figure}[!htp]
    \centering
\includegraphics[height=5.5cm,width=8cm]{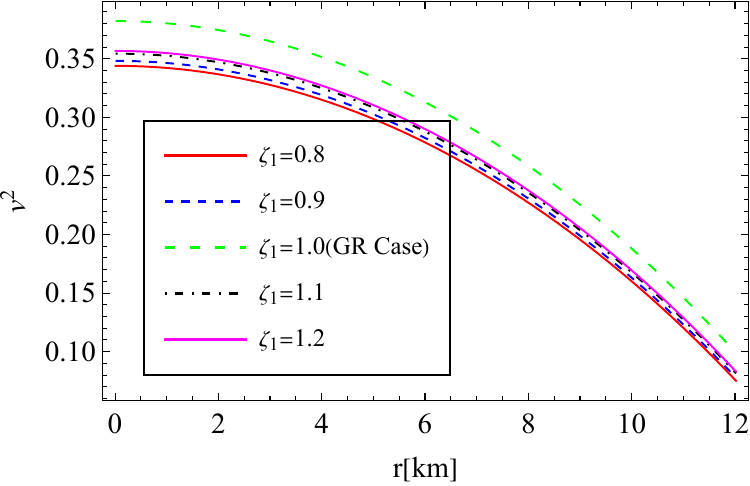}
    \caption{Profile of velocity of sound ($v^2$) versus `r'. }
    \label{fig5}
\end{figure}
The velocity of sound waves in the charged isotropic fluid must be slower than electromagnetic radiation (light travels at the speed of light, $c = 1$). On using Herrera's over-tuning technique~\cite{Abreu:2007ew} for our model's stability, which specifies that the interval's region should be $0 < v^2 = \frac{d\mathcal{P}}{d\rho} < 1$.  The mathematical formula for the velocity of sound are as follows:\\

\begin{equation}\label{eq46}
    v^2 = \frac{d\mathcal{P}}{d\rho}
\end{equation}
The expression of the velocity of sound for current analysis is defined as
\begin{eqnarray}\label{eq47}
&& \hspace{-0.5cm}   \frac{d\mathcal{P}}{d\rho}=\frac{\eta_{8}(r)-\eta_{9}(r)-\frac{\lambda(\mu+\lambda r^2) \eta_{10}(r) \eta_{11}(r)}{\eta_{12}(r)+\eta_{24}(r)\eta_{25}(r)}}{\eta_{22}(r)+\eta_{23}(r)}
\end{eqnarray}

\begin{figure}[!htp]
    \centering
\includegraphics[height=6cm,width=7.5cm]{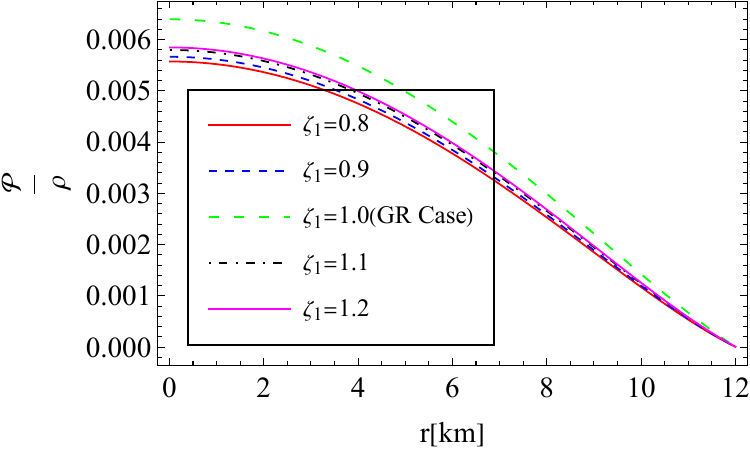}
    \caption{Profile of ratio of pressure and density versus `r'}
    \label{fig6}
\end{figure}
 Fig.~\ref{fig5} illustrate the velocity of sound across the star is slower than the speed of light. Additionally, we have examined the validity of the charged matter and if dark matter can be presumed from the assured restriction that applies to the EoS components, such as $0<w< 1$. These matter-related limitations must be followed to ensure the realistic composition of charged and uncharged matter. The EoS mathematical expression is as follows:
 \begin{equation}\label{eq48}
     w=\frac{\mathcal{P}}{\rho}
 \end{equation}
 It is simple to conclude that the realistic character of matter composition under EoS is predicted by our system of celestial bodies. The graphs for the profile $ w=\frac{\mathcal{P}}{\rho}$ is displayed in Fig.~\ref{fig6}.
 \subsection{Adiabatic Index}
 For the first time, Chandrasekhar addressed the stability of celestial structures under adiabatic index $\Gamma$ in his works~\cite{Chandrasekhar:1964zz,Chandrasekhar:1964zza}. Many of the contributors~\cite{Bowers:1974tgi, Andreasson:2007ck} highlighted this stability limit in their discussion. Heintzmann and Hillebrandt~\cite{Heintz} expanded on this stability approach in their work by adding an acceptability constraint, such as $\Gamma_{|0\leq r\leq R} > \frac{4}{3}$. The adiabatic index is expressed mathematically as
 \begin{figure}[!htp]
    \centering
\includegraphics[height=6cm,width=7.5cm]{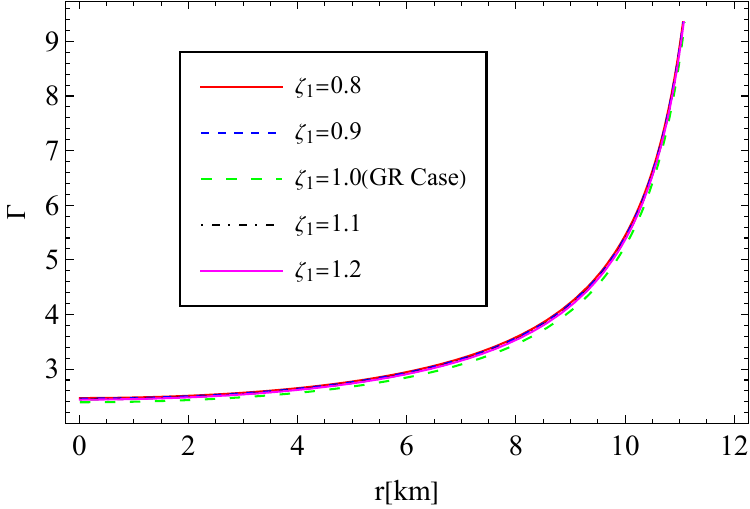}
    \caption{Profile of adiabatic index~($\Gamma$) versus `r'.}
    \label{fig7}
\end{figure}
 \begin{equation}\label{eq49}
     \Gamma =\Bigg[ 1+\frac{\rho}{\mathcal{P}}\Bigg]~\frac{d\mathcal{P}}{d\rho}
 \end{equation}
 The stability of our electrically powered stellar system under adiabatic index requirements is displayed clearly in the Fig.~\ref{fig7}. According to Fig.~\ref{fig7}, the adiabatic index looks to have a low value at the center and to be increasing. The value of $\Gamma$ at the core with constant $\zeta_2 =0.00002 km^{-2}$ is as follows: (i) $\Gamma_{0}=2.47$ for $\zeta_{1}=0.8 $, (ii)  $\Gamma_{0}=2.46$ for $\zeta_{1}=0.9 $, (iii)  $\Gamma_{0}=2.41$ for $\zeta_{1}= 1.1$, (iv) $\Gamma_{0}=2.43$ for $\zeta_{1}= 1.2$, \& with $\zeta_2=0$, the value $\Gamma_{0}=2.39$ for $\zeta_1 =1.00$ ~(See Table~\ref{tab2}). The values of $\Gamma$ show that it is greater than 4/3, proving that a stable, ideal fluid-charged model has been achieved.

 \section{Modified Buchdahl limit of gravitational collapse in $\mathscr{F}(Q)$-gravity}
 \label{sec: Modified. Buchdhal.}

In order to assess the compactness of a stellar object, it is essential to examine the upper limit of the mass-radius ratio. \cite{Buchdahal} initially proposed the concept of the maximum mass-radius ratio in the setting of a perfect fluid with a decreasing energy density approaching the surface. This scenario accurately captures the distribution of isotropic matter. The upper limit on mass-radius can be expressed as follows: 
 \begin{eqnarray} \label{eq77}
&&\hspace{1cm}\frac{\tilde{\mathcal{M}}}{R}\leq\frac{4}{9}.
\end{eqnarray}
The symbol $\tilde{\mathcal{M}}$ denotes the overall mass of an object in a uniform distribution of matter, whereas $R$ denotes the radius of the object, which is derived by considering zero pressure at the object's surface. 
In addition, the existence of an electric charge in the solution modifies Buchdahl's limit. The lower and upper bounds on the mass-radius ratio in this situation were provided by \cite{Andr} and \cite{Bohmer}, respectively, within the context of the standard GR theory.  
Hence, the constraints on the mass-radius ratio within the framework of General Relativity (GR) may be expressed as follows:  
\begin{eqnarray}
\label{eq78}
\frac{Q^2\left[18R^2+Q^2\right]}{2R^2\left[12R^2+Q^2\right]}\leq \frac{\mathcal{M}}{R}\leq \frac{1}{9R^2}\Big[2R\sqrt{3Q^2+R^2}+3Q^2+2R^2\Big].~~
\end{eqnarray}
Moreover, \cite{Bohmer} also derived an overall formula for the smallest possible value of the mass-radius~(M-R) relationship by including a cosmological constant which can be given by~, 
\begin{eqnarray}
    \frac{9 Q^2}{12R^2+Q^2}  \bigg[ \frac{9 Q^2+\Lambda R^4}{9 Q^2} +\frac{Q^2}{18 R^2} -\frac{\Lambda R^2}{54}  \bigg] \le \frac{\mathcal{M}}{R}.~~ \label{eq79}
\end{eqnarray}
It is important to mention that when $\Lambda=0$, we get the minimum value for the M-R ratio as estimated in GR.  The Eq. (\ref{eq79}) is expanded using the binomial expansion, which allows us to get
\begin{eqnarray}
  \bigg[\frac{3Q^2}{4R^2}+\frac{\Lambda R^2}{12}-\frac{9 \Lambda Q^2}{432} + O \bigg(\frac{Q^4}{R^4}\bigg) \bigg] \le \frac{\mathcal{M}}{R}.  \label{eq80}
\end{eqnarray}
Given that $\frac{Q^2}{R^2}\ll 1$ and $\Lambda Q^2\ll 1$ in our current scenario, then we can confidently disregard the terms $O(\frac{Q^4}{R^4})$ and $\Lambda Q^2$ in Eq.(\ref{eq80}). Therefore, we can now get the most accurate equation for the minimum value of the mass-radius relation 

\begin{eqnarray}
\bigg[\frac{3Q^2}{4R^2}+\frac{\Lambda R^2}{12} \bigg] \le  \frac{\mathcal{M}}{R} . \label{eq81} 
\end{eqnarray}

Furthermore, \cite{Andreasson:2012dj} calculated the maximum value for the M-R relationship for the charged fluid configuration by taking a positive cosmological constant value as
\begin{eqnarray}
    \frac{\mathcal{M}}{R} \le \frac{2}{9}+\frac{Q^2}{3R^2}-\frac{\Lambda R^2}{9}+\frac{2}{9} \sqrt{1+\frac{3Q^2}{R^2}+3 \Lambda R^2}. \label{eq82}
\end{eqnarray}
When the parameter $\Lambda$ is equal to zero in Eq. (\ref{eq82}), it is evident that the maximum limit of $\frac{\mathcal{M}}{R}$ is restored in the context of GR. Nevertheless, in our particular situation, we have determined that $\Lambda$ is equal to $\frac{\zeta_2}{2\zeta_1}$ and it is more than zero. Hence, the bounds of M-R relation in $\mathscr{F}(Q)$ -gravity theory that satisfies the outer spacetime (\ref{eq43}) may be precisely defined via the subsequent inequality:

\begin{eqnarray}
   \bigg(\frac{3Q^2}{4R^2}+\frac{\zeta_2 R^2}{24 \zeta_1} \bigg) \le  \frac{\mathcal{M}}{R} \le \bigg( \frac{2}{9}+\frac{Q^2}{3R^2}-\frac{\zeta_2 R^2}{18\zeta_1}+\frac{2}{9} \sqrt{1+\frac{3Q^2}{R^2}+\frac{3\zeta_2 R^2}{2 \zeta_1}} \bigg),~~~\label{eq83}
\end{eqnarray}
Here $\mathcal{M}$ denotes the total mass of the charged compact star.  It is important to emphasize that the mass represented by $\mathcal{M}$ in Eqs. (\ref{eq82}) or (\ref{eq83}) does not correspond to the total mass ($\hat{\mathcal{M}}$) specified in Eq.(\ref{eq77}). The formula for $\mathcal{M}$ may be derived in the described form: 
\begin{eqnarray}
\mathcal{M}  = \frac{R}{2}\bigg[1-e^{-\chi(R)}-\frac{\Lambda}{3}R^2+\frac{Q^2}{R^{2}}\bigg]=\frac{R}{2}\bigg[1-e^{-\chi(R)}-\frac{\zeta_2}{6\zeta_1}R^2+\frac{q^{2}(R)}{R^{2}}\bigg].~~~\label{eq84}  
\end{eqnarray}
After substituting of the values of $e^{-\chi(R)}$ and $Q$, we get,
\begin{eqnarray}
&&\hspace{-0.5cm}  \mathcal{M}(R)=  \frac{R}{2}\Bigg[1-\frac{\mu+\lambda R^2}{\mu(1+\lambda R^2)} -\frac{R^2 \zeta_2}{6 \zeta_1}-\frac{\zeta_1 \lambda^4 R^2 \Big(\mu+\frac{2 \varpi ^2 \left(\eta^2+1\right)}{\varpi ^2 \eta^2+\eta}-\frac{3 \eta^2-2}{4 \left(\eta^2+1\right)}-1\Big)}{2 \mu \left(\lambda R^2+1\right)^2}\Bigg]. 
\end{eqnarray}
\begin{figure}[!htp]
    \centering
\includegraphics[height=5cm,width=7cm]{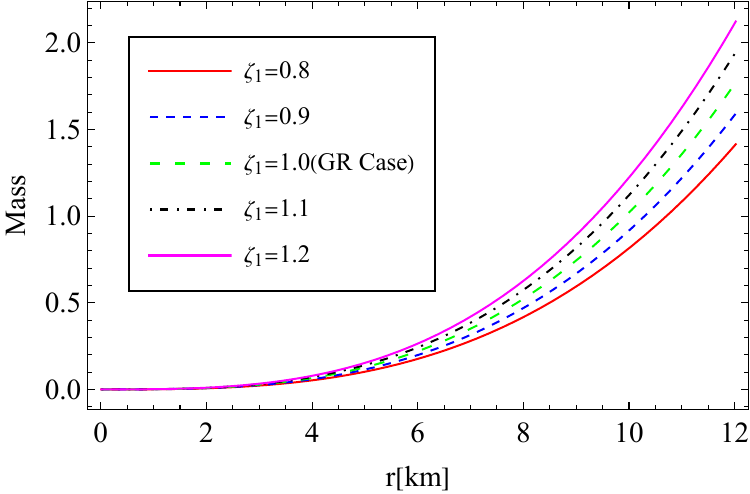}
    \caption{Profile of mass function versus `r'.}
    \label{fig8}
\end{figure}

Nevertheless, when dealing with a perfect fluid or anisotropic fluid matter distribution, the gravitational mass (as represented by Eq.~\ref{eq84}) is indistinguishable from the effective mass. However, this does not apply to distributions of charged objects. In the setting of $\mathscr{F}(Q)$ -gravity, the effective mass for the charged object may be calculated using the following equation: 
\begin{eqnarray}
\mathcal{M}^{\textrm{eff}}=\frac{1}{2 \zeta_1}\int_{0}^{R}\left({\rho}+\frac{q^{2}}{r^4}+\frac{\zeta_2}{2}\right)r^{2}dr =\frac{R}{2}\left[1-e^{-\chi(R)}\right].~~ \label{eq85}
\end{eqnarray}
After doing a thorough analysis of Eq. (\ref{eq85}), the following conclusions may be drawn: (i) In $\mathscr{F}(Q)$-gravity, the total mass $\mathcal{M}$ in Eq.(\ref{eq84}) is greater than the total mass $\hat{\mathcal{M}}$ given by Eq.(\ref{eq77}). (ii) The effective mass $\mathcal{M}_{\textrm{eff}}$ of the charged compact star model is equal to the total effective mass for perfect fluid or anisotropic fluid matter distribution. \\

 It has been observed in many astronomical situations, such as those close to black holes and those with massive galaxies. Finally, a key component of modern physics that has been extensively studied and seen is the gravitational redshift. It has a profound impact on our understanding of gravity and the universe in general. It is significant to remember that the interior redshift is lowest at the surface and highest in the vicinity of the core. More precisely, the gravitational redshift of the compact star is given by the formula:
 \begin{equation}\label{eq50}
     Z_g= \sqrt{e^{-\chi(r)}}-1
 \end{equation}
 while the surface redshift is given by,
 \begin{equation}\label{eq51}
     Z_s =(1-2u)^{-\frac{1}{2}}= \sqrt{e^{{\Upsilon(R)}}}-1  
 \end{equation}
 where $u=\frac{\mathcal{M}^{\textrm{eff}}}{R}$.
 \begin{figure}[!htp]
    \centering
\includegraphics[height=5cm,width=7cm]{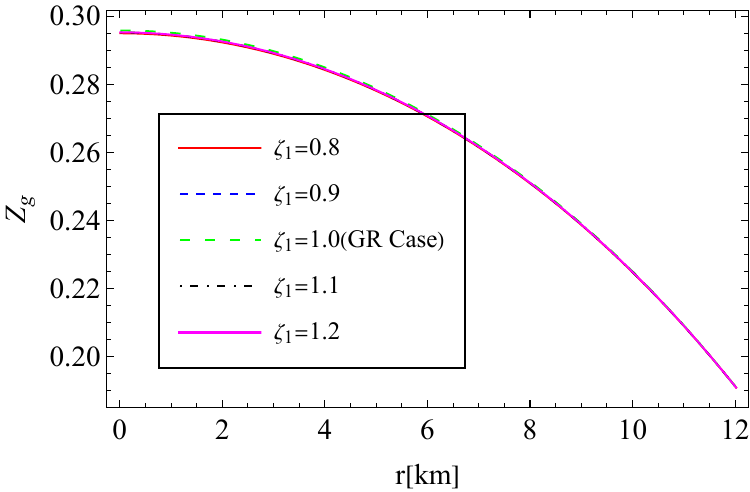}
    \caption{Profile of  $Z_g$ versus `r'}
    \label{fig9}
\end{figure}
 It has been confirmed~\cite{Buchdahl:1959zz,Straumann:1984xf} that the surface redshift for spherically symmetric ideal fluid spheres is $Z_s < 2$, and in the case of anisotropic cases, this value comes out to be 3.84, as in~\cite{Karmakar:2007fn,Barraco:2003jq}. However, as demonstrated by Boehmer and Harko~\cite{Boehmer:2006ye}, this number could be reached to $Z_s \le 5,211$, which is in line with the bound $Z_s \leq 5.211$ found by Ivanov~\cite{Ivanov:2002jy}. Fig.~\ref{fig9} makes it very evident that the compact object behaves naturally and that the redshift values are well within Ivanov's suggested bounds.\\
 A graph representing the compact star's mass can be found in Fig.~\ref{fig8}. The effect of $\zeta_1$ on the mass of the compact star is readily seen. A rise in the value of $\zeta_1$ corresponds with a mass value. The weird star's maximum compactness limit was determined to be $\frac{M}{R}<\frac{4}{9}$~\cite{Bowers:1974tgi}.



\begin{table}
\centering
\caption{The numerical data of central density~($\rho_c$), surface density~($\rho_s$), central pressure~($\mathcal{P}_c$) and M-R ratio at the center for the different values of $\zeta_1$}
\begin{tabular}{ccccc}
\hline$\zeta_1$ & $\rho_c~(gm/cm^3)$ & $\rho_s~(gm/cm^3)$ & $\mathcal{P}_c~(dyne/cm^2)$ & $\Gamma(r=0)$ \\
\hline 0.8 & $4.15579 \times 10^{15}$ & $4.04041 \times 10^{15}$ & $2.08495\times 10^{34}$ & 2.47 \\
0.9 & $4.67500 \times 10^{15}$ & $4.54680 \times 10^{15}$ & $2.38500 \times 10^{34}$ &  2.46\\
1.0 & $5.20489 \times 10^{15}$ & $5.06174 \times 10^{15}$ & $2.99664 \times 10^{34}$ & 2.39 \\
1.1 & $5.71769 \times 10^{15}$ & $5.55958\times 10^{15}$ & $2.98317 \times 10^{34}$ & 2.41 \\
1.2 & $6.23690 \times 10^{15}$ & $6.06383 \times 10^{15}$ & $3.28322 \times 10^{34}$ & 2.43 \\
\hline
\end{tabular}

\label{tab2}
\end{table}
With a minimum value of $\zeta_1$, the star obtained of 1.3$M_{\odot}$ and falls within the range of stars that can be considered compact stars. The graph's growing trend suggests that when the value of $\zeta_1$ rises, the star's mass increases gradually. The mass function value is not significantly affected by the constant $\zeta_2$, since its value is assumed to be 0.00002 in the current investigation.\\

In Table~\ref{tab3}, the notations $L_Q$ and $U_Q$ denotes the lower and upper bound in $\mathscr{F}(Q)$ gravity theory, which are defined as 
\begin{eqnarray}
    L_Q=\Big(\frac{3Q^2}{R^2}+\frac{\zeta_2 R^2}{24\zeta_1}\Big),~~~ U_{Q}=\left(\frac{2}{9}+\frac{{Q}^2}{3 R^2}-\frac{\zeta_2 R^2}{18 \zeta_1}+\frac{2}{9} \sqrt{1+\frac{3 Q^2}{R^2}+\frac{3 \zeta_2 R^2}{2 \zeta_1}}\right) .
\end{eqnarray}


\begin{table*}[!htp]
 \caption{ The numerical values of lower bound~($L_Q$), M-R ratio, upper bound~($U_Q$), and surface red-shift~($ Z_s$) of the star for different parameter values of $\zeta_1$ and $\zeta_2$ .}
\begin{tabular}{c|c|c|c|c}
\hline Parameter & $L_Q$-$\mathscr{F}(Q)$ gravity & $M/R$ & $U_Q$-$\mathscr{F}(Q)$ gravity & $Z_s$. \\
\hline$\zeta_1=0.8, \zeta_2=0.00002$ & 0.0415762 & 0.156901 & 0.480918& 0.190858 \\
$\zeta_1=0.9, \zeta_2=0.00002$ & 0.0525632 & 0.158157 & 0.490248 &  0.190858 \\
$\zeta_1=1.0, \zeta_2=0.00000$ & 0.0647276 & 0.159646 & 0.500326 &  0.190858 \\
$\zeta_1=1.1, \zeta_2=0.00002$ & 0.0784299 & 0.160649 & 0.511930 &  0.190858 \\
$\zeta_{1}=1.2, \zeta_2=0.00002$ & 0.0933081 & 0.161890 & 0.524229 &  0.190858\\
\hline
\end{tabular}
  \label{tab3} 
\end{table*}

\section{Eqilibrium analysis via  Tolman-Oppenheimer-Volkoff~(TOV) equation}
\label{sec: TOV.}
The TOV equation, which is the radial component of Einstein's field equations for a spherically symmetric space-time in GR, is presented below. The equilibrium equation for a static, isotropic and spherically symmetric fluid star in GR are given by this equation~\cite{Tolman:1939jz,Maurya:2017uyk}.
\begin{figure}[!htp]
    \centering
\includegraphics[height=8cm,width=12cm]{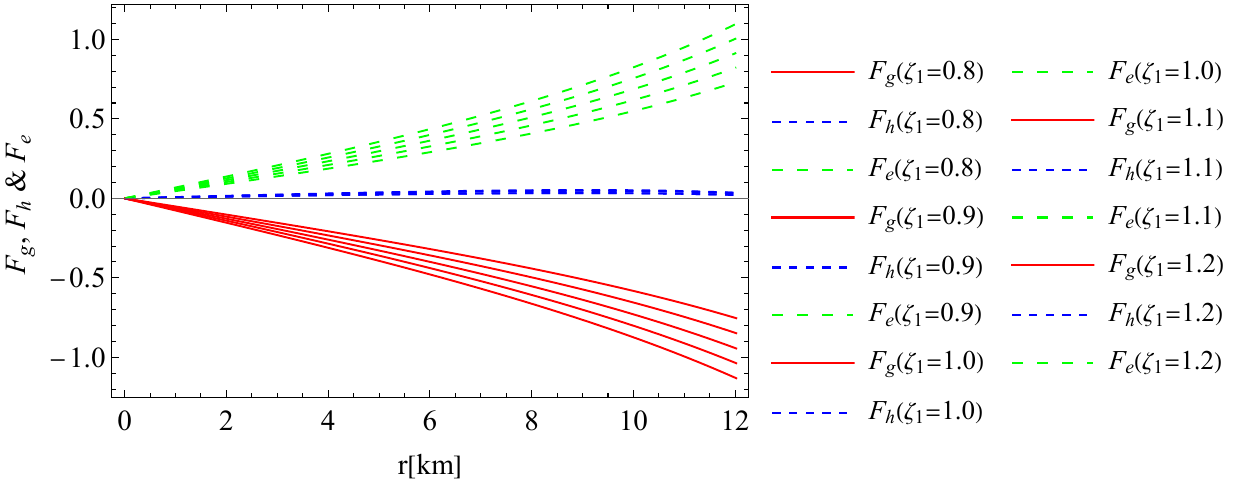}
    \caption{Profile of different forces~($F_h$, $F_e$ and $F_g$) versus `r'.}
    \label{fig10}
\end{figure}
\begin{equation}\label{eq54}
-\frac{d \mathcal{P}}{d r}-\frac{\chi^{\prime}}{2}\left(\mathcal{P}+\rho\right)+\sigma(r) E(r) e^{\frac{\chi(r)}{2}}=0,
\end{equation}
Now let us try to interpret the preceding equation from the perspective of equilibrium, where the expression is viewed as a composition of three separate forces: electric ($F_e$), hydrostatic ($F_h$), and gravitational ($F_g$). The following forces apply to our system:
\begin{equation}\label{eq55}
\begin{aligned}
& F_g=-\frac{\chi^{\prime}}{2}(\rho+\mathcal{P}), \\
& F_h=-\frac{d \mathcal{P}}{d r}, \\
& F_e=\sigma \frac{q}{r^2} e^{\frac{\lambda}{2}}=\frac{1}{8 \pi r^4} \frac{d q^2}{d r},
\end{aligned}
\end{equation}
where $
\text { where } F_g+F_e+F_h=0 \text {. }
$\\

Fig.~\ref{fig10} shows a graphical depiction of all these various forces for different $\zeta_1$ values. A compact star needs all of the previously listed forces to equal zero to be stable. Fig.~\ref{fig10} makes this quite evident: the gravitational force ($F_g$) acts inward, suggesting that it is attracting. The hydrostatic and electrostatic forces act outward and are repulsive ($F_h$ and $F_e$ respectively). Therefore, it is accurate to state that the repulsion of hydrostatic and electrostatic forces balances the attractive gravitational forces, stabilizes the forces surrounding the star, and provides us with a stable, compact star that is in perfect equilibrium.
\section{Investigationsof the mass-radius relation of observed compact objects  using M-R curve}\label{sec: MR Curve}
\begin{figure*}[!htp]
    \centering
\includegraphics[height=8cm,width=10cm]{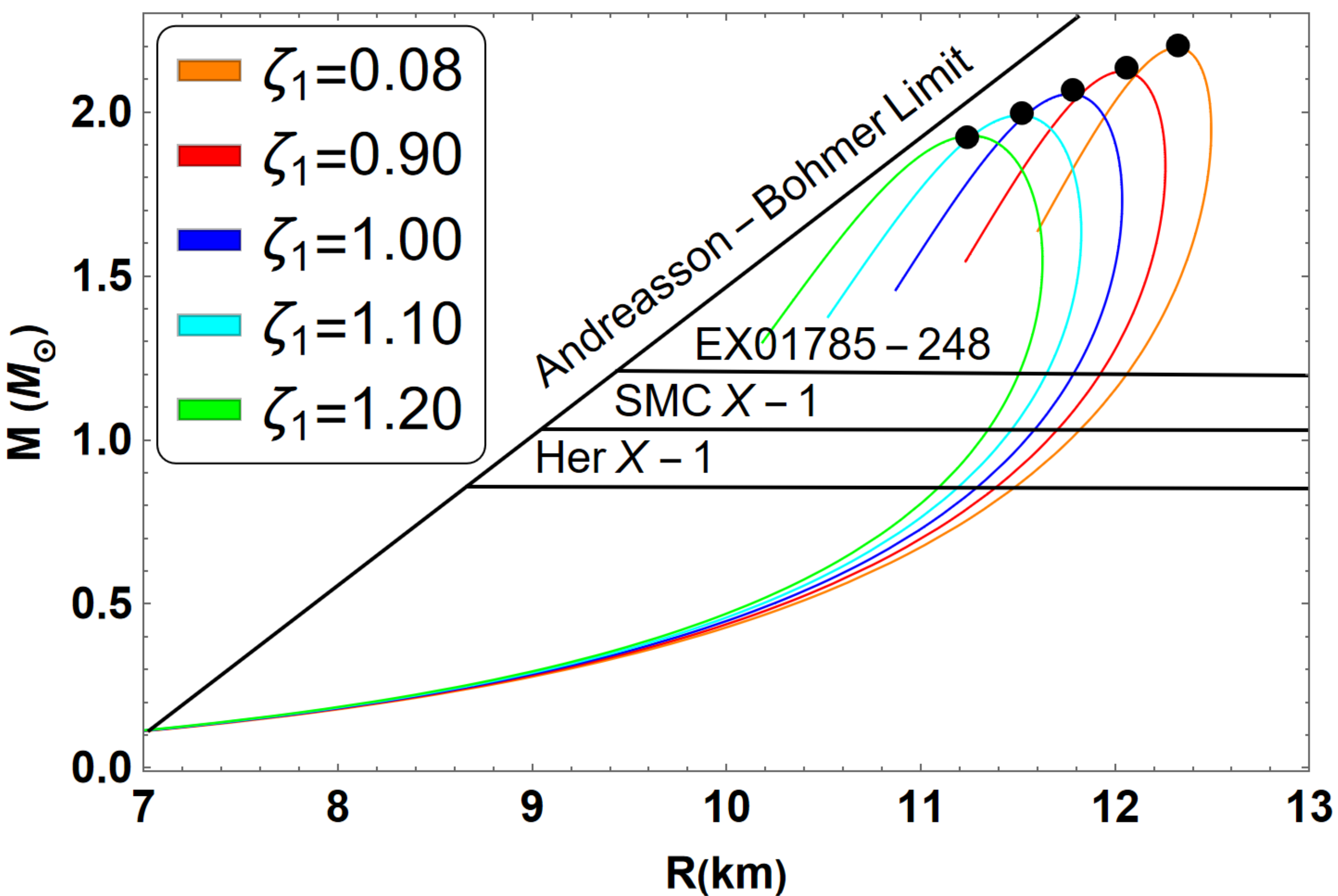}`
    \caption{The graphical evolution of $M-R$ with $\mu=0.0000294  $, $\lambda=-0.00000006 $, $\varpi=7.368853$ for different values of $\zeta_1$ }
    \label{fig11}
\end{figure*}

This section examines the relationship between mass and radii, which is a crucial study for the physical acceptability of the current model. In this regard, we plot the M-R curve behavior for various values of $\zeta_1$ as shown in Fig.~\ref{fig11}. We provide three stellar candidates in Fig~\ref{fig11} with known measured masses and estimated their radii. With the different values $\zeta_1$ with fixed $\zeta_2$, the estimated radii are listed in Table~\ref{tab5}. The maximum mass reduces as the value of the coupling parameter $\zeta_1$ becomes larger, as shown in Fig.~\ref{fig11}. Three distinct star candidates were selected from the literature: Her X-1~\cite{Abubekerov:2008inw} with an associated mass of 0.7-1~$M_\odot$, SMC X-1~\cite{Rawls:2011jw} with an associated mass of 0.95-1.13~$M_\odot$, and EX01785-248~\cite{Ozel:2008kb} with an associated mass of 1.1-1.5~$M_\odot$.\\

\begin{table}[!htp]
\caption{The numerical values of maximum mass and maximum radius via M-R curve.}
\centering
\begin{tabular}{c|c|c} 
\hline
Parameters & Maximum Mass~($M_\odot$)  & Maximum Radius~(km) \\
\hline
$\zeta_1=0.8~(\mystar[orange] )$ & 2.321 & 12.09 \\
$\zeta_1=0.9~(\mystar[red] )$ & 2.128  & 12.00 \\
$\zeta_1=1.0~(\mystar[blue] )$ & 2.042 & 11.79 \\
$\zeta_1=1.1~(\mystar[cyan] )$ &1.989 & 11.51 \\
$\zeta_1=1.2~(\mystar[green] )$ & 1.927 & 11.25 \\
\hline
\hline
\end{tabular}

\label{tab4}
\end{table}

Our analysis presented graphically, shows that we have predicted the above mass from our current model for a range of values of the coupling parameter $\zeta_1$ related to the $\mathscr{F}(Q)$ gravity. The $\mathscr{F}(Q)$ theory of gravity has the ability to increase maximum masses, enabling the stellar model to overcome gravitational collapse with a greater mass. The gravitational wave GW190814, a binary merger of two compact objects, was identified by LIGO/VIRGO experiments as the first ever ``mystery object'' with a mass of 2.50-2.67~$M_\odot$ and a black hole mass of 22.2–24.3~$M_\odot$~\cite{LIGOScientific:2020zkf}. Within the ``mass gap'' created by cosmic collisions, the object is either the lightest black hole or the heaviest neutron star. With the current model, we have calculated the maximum allowable mass of 2.321~$M_\odot$, which is attained for $\zeta_1=0.8$~(See Table~\ref{tab5}) and falls within the range suggested by the LIGO/VIRGO experiment~\cite{LIGOScientific:2020zkf}.

\begin{table}[!htp]
\caption{{Predicted radii of different stellar candidates from the model for $\mathscr{F}(Q)$ gravity}}
\centering
\begin{tabular}{|c|c|c|c|c|c|c|c|c|}
    \hline
    & \multicolumn{6}{|c|}{Predicted radii (km) } \\
    \hline
Stellar objects\;\;\;\;                       & ${\frac{M}{M_\odot}}$\;\;\;\       &$\zeta_1=0.80$\;\;\;     &$\zeta_1=0.90$     &$\zeta_1=1.00$     &$\zeta_1=1.10$     &$\zeta_1=1.20$ \\
\hline
Her X-1~\cite{Abubekerov:2008inw}                               &0.85\;                              &11.08                  &11.19                        &11.28                &11.38                         &11.48\\
SMC X-1~\cite{Rawls:2011jw}                               &1.04\;                              &11.35                  &11.49                        &11.60                 &11.72                       &11.86\\
EX01785-248~\cite{Ozel:2008kb}                             &1.30\;                              &11.57                &11.74                        &11.87                 &12.02                       &12.18\\

\hline
\end{tabular}
\label{tab5}
\end{table}

\section{Concluding Remarks}
\label{sec: Conclusion}
For several reasons, compact objects are important in relativistic astrophysics. Neutron stars are among the universe's most stable compact objects; however, the maximum mass of these objects remains unknown to scientists. In this study, we have considered the Buchdahl ansatz~\cite{Buchdahl:1959zz} to exactly address the stability limit of relativistic charged spheres in the context of Einstein-Maxwell theory.  This would especially apply to the electric fields generated by assumed compact stars composed of ideal fluid matter, where the Reissner-Nordstrom metric is used to represent the external spacetime. Our method is novel for the reason that we interpret the system of hypergeometric equations by first building the charged fluid sphere from Buchdahl ansatz through the appropriate transformation. Identification of the stellar structure is largely dependent on the parameter $\mu$. This arrangement guarantees a stellar structure free of singularities since the energy density and pressure are maximal in the center and monotonically decrease towards the outer limits. It is noteworthy the behavior of a compact star, as defined by $\mathscr{F}(Q)$ gravity, is consistent with GR.
 Consequently, the sources of $\mathscr{F}(Q)$ gravity can be used to derive any solution found in general relativity. We generate a new solution in the $\mathscr{F}(Q)$ gravity theory using the well-behaved Buchdahl ansatz in order to obtain an obedient solution.\\
 
 (i) The problem has undergone examination of the model's thermodynamic quantities, and the findings are presented in the figures. In addition, Table~\ref{tab2} provides the mathematical values of the thermodynamic quantities, Fig.~\ref{fig2} illustrates how the model's density,  pressure, and charge behave about the constant $\lambda =-0.00000006$. It should be mentioned that the values of these thermodynamic quantities depend on the parameter $\zeta_1$, with 0.8, 0.9, 1.0, 1.1, and 1.2. We observe a consistent rise in all quantities as the values of $\zeta_1$ rise. Every graph in the illustration exhibits good behavior, and every regularity criterion is met. The numerical values of central density~($\rho_c$), surface density~($\rho_s$) and central pressure~($\mathcal{P}_{c}$) can be observed in Table~\ref{tab2}.\\

 (ii) The energy conditions of a compact star are studied in Fig.~\ref{fig3}. All of the energy conditions, which guarantee that SEC, NEC, and WEC are given to be true, are satisfied for the compact star, as evidenced by the positive behavior of the thermodynamic quantities in Fig.~\ref{fig2}. As seen in the above graphic, we plotted the dominating energy conditions graph. It is shown that the compact star is well-behaved. Within the neighborhood of the compact star, the values of SEC and WEC are positive.\\

 (iii) The compact star's adiabatic index, mass, and redshift are depicted in Figs.~\ref{fig7}, \ref{fig8} \& \ref{fig9}. In this case, the rise in $\zeta_1$ values likewise corresponds to an increase in the compact star's mass. According to the Buchdahl limit and the study by Ivanov~\cite{Ivanov:2002jy,Buchdahl:1959zz}, this limit suggests that $Z_s\leq4.77$ for the compact star and that the graph's trajectory is decreasing monotonically throughout the star's vicinity. The adiabatic index has an increasing trajectory as the value of $\Gamma>4/3$ and the mass \& gravitational redshift of compact stars are well-behaved.\\

 (iv) A graphic depiction of the various forces in the compact star model is shown in Fig.~\ref{fig10}. The graph demonstrates the fulfillment of the TOV equation. The visual depicts the gravitational force ($F_g$), which acts inward and is attractive. On the other hand, the outward behavior of certain forces, including the hydrostatic force ($F_h$) and electrostatic force ($F_e$), indicates their repulsive nature. The sum of all the forces is zero i.e. $F_e + F_g + F_h = 0$.\\
 
(v) We verified that, for the current model, the maximum masses and matching radii have risen as the coupling parameter $\zeta_1$ connected to the $\mathscr{F}(Q)$  gravity has decreased. The recent analysis of the M-R curve done by Tangphati et al~\cite{Tangphati:2022cqr}, which shows that the maximum mass and radius rise when the Rastall parameter $\eta$ decreases. With a radius of 11.77 km, they arrive at a maximum mass of 2.364~$M_\odot$. The study conducted by Deb et al.~\cite{Deb:2018sgt} demonstrates that the star systems under examination steadily increase in mass and size as the value of $\chi$, the coupling parameter of $f (R, T)$ gravity, decreased. Furthermore, our suggested model was able to accurately forecast the masses of a few huge star objects, such as the Her X-1~\cite{Abubekerov:2008inw} with an associated mass of 0.7-1~$M_\odot$, SMC X-1~\cite{Rawls:2011jw} with an associated mass of 0.95-1.13~$M_\odot$, and EX01785-248~\cite{Ozel:2008kb} with an associated mass of 1.1-1.5~$M_\odot$. The finding satisfies the observational restrictions and may be a contender for the lighter object in GW190814~\cite{LIGOScientific:2020zkf} and the huge unseen companion in the binary system 2MASS J05215658+4359220 with mass $3.3^{+2.8 }_{−0.7}$~\cite{Thompson:2018ycv}. A model of a quark star in a non-minimal geometry-matter coupling theory of gravity was recently proposed by Carvalho et al.~\cite{Carvalho:2022kxq}. They also achieved a mass of 2.6~$M_\odot$ by using $\sigma = 50 km^2$. The highest mass that we were able to get agrees with the findings of Carvalho et al.~\cite{Carvalho:2022kxq}. After conducting our analysis, we have concluded that this gravity theory is able to generate large dark energy stars and offer a realistic explanation of the M-R relation that meets the constraints of observation. \\

 To sum up, this research offers a framework for studying isotropic matter distribution in the presence of electric charge using the metric potential provided by Buchdhal in $\mathscr{F}(Q)$ gravity. Observe that the formulation of $\mathscr{F}(Q)$ gravity produces identical outcomes to those of general relativity, suggesting that the sources of $\mathscr{F}(Q)$ gravity can be utilized to obtain any solution found in general relativity. The $\mathscr{F}(Q)$ gravity theory offered a new solution when we applied the Buchdhal potential to get a well-behaved solution. In conclusion, the special coordinate transformation and $\mathscr{F}(Q)$ gravity together provide a practical way to produce charged isotropic solutions in both modified gravity theory and general relativity.
 Additionally, we point out that our work has physical validity and might help with future research in the framework of $\mathscr{F}(Q)$ -gravity theory with other well-known metric potentials.

\appendix
\section{Appendix}
\subsection{Gupta-Jasim Two Step Method}
  For the Eq.~(\ref{eq38}), we compare it with standard form $2^{nd}$ order differential equation 
  $$
W_0 \frac{d^2 \mathcal{W}}{d \eta^2}+W_1 \frac{d \mathcal{W}}{d \eta}+W_{2}~\eta=R
$$
where $W_0=\eta^2 \varpi^2+\eta, W_1=0, W_2=-2 \eta^2, R=0$. To obtain exactness, one can verify that $W_2-\frac{d W_1}{d \eta}+\frac{d^2 W_0}{d \eta^2}=0$, which shows that the equation Eq.~(\ref{eq38}) is exact. Thus, the primitive of the given differential equation is
$$
\begin{gathered}
W_0 \frac{d \mathcal{W}}{d\eta }+\left(W_1-W_0^{\prime}\right) \eta=\int R d \eta+A, \\
\frac{d \mathcal{W}}{d \eta}-\frac{2 \eta \varpi^2+1}{\eta^2 \varpi^2+\eta} \mathcal{W}=\frac{A}{\eta^2 \varpi^2+\eta}
\end{gathered}
$$
where $A$ is an arbitrary constant. On solving the above equation, which yields the solution.
\begin{eqnarray*}
&& \hspace{-0.5cm}\mathcal{W}(\eta)=A \Big[-2 \varpi ^4 \eta^2 \log (\eta)+2 \varpi ^4 \eta^2 \log \left(\varpi ^2 \eta+1\right)\nonumber\\&&\hspace{0.5cm}-2 \varpi ^2 \eta-2 \varpi ^2 \eta \log (\eta)+2 \varpi ^2 \eta \log \left(\varpi ^2 \eta+1\right)-1\Big]\nonumber\\&&\hspace{4.5cm}+B~ \eta \left(\varpi ^2 \eta+1\right),
\end{eqnarray*}
where A and B are the integration constants.
\subsection{The expressions  $\eta_{8}$ to $\eta_{28}$ appears in present work:}
\begin{small}
\begin{eqnarray*}
&& \hspace{-0.5cm}\eta_{8}=\frac{\zeta_1 \lambda^2 r^2 \Big[\frac{4 \varpi ^2 \lambda r}{(1-\mu) \left(\eta-\varpi^2 \eta^2\right)}-\frac{2 \varpi ^2 \left(\lambda r^2+1\right) \left(\frac{\lambda r}{(1-\mu) \eta}-\frac{2 \varpi ^2 \lambda r}{\mu-1}\right)}{(1-\mu) \left(\eta-\varpi^2\eta \right)^2}-\frac{8 \lambda r \left(-3 \lambda r^2-5 \mu+2\right)}{\left(4 \lambda r^2+4\right)^2}-\frac{6 \lambda r}{4 \lambda r^2+4}\Big]}{2 \mu \left(\lambda r^2+1\right)^2}+\frac{2 \zeta_1 \lambda^2 (\mu-1) \mu r}{\left(\lambda \mu r^2+\mu\right)^2},\\
&& \hspace{-0.5cm}\eta_{9}=\frac{2 \zeta_1 \lambda^3 r^3 \Big[\frac{2 \varpi ^2 \left(\lambda r^2+1\right)}{(1-\mu) \left(\eta-\varpi ^2\eta^2 \right)}+\frac{-3 \lambda r^2-5 \mu+2}{4 \lambda r^2+4}+\mu-1\Big]}{\mu \left(\lambda r^2+1\right)^3}+\frac{\zeta_1 \lambda^2 r \Big[\frac{2 \varpi ^2 \left(\lambda r^2+1\right)}{(1-\mu) \left(\eta-\varpi ^2 \eta^2\right)}+\frac{-3 \lambda r^2-5 \mu+2}{4 \lambda r^2+4}+\mu-1\Big]}{\mu \left(\lambda r^2+1\right)^2},
\end{eqnarray*}
\begin{eqnarray*}
&& \hspace{-0.5cm}\eta_{10}=-8\varpi^2 A-\varpi^2 A \left(\lambda r^2 \left(5 \varpi ^2 \eta+3\right)+\varpi ^2 K \eta+4 \varpi ^2 \eta+\mu+2\right) \log \left(\eta\right)+2 \varpi ^2 A \Big(\lambda r^2 \left(5 \varpi ^2 \eta+3\right)+\varpi ^2 \mu \eta \nonumber\\ &&\nonumber\\ && \hspace{0.5cm}+4 \varpi ^2 \eta+K+2\Big) \log \left(\varpi ^2 \eta+1\right)-A \eta+A \mu \eta-10 \varpi ^2 A \lambda r^2-2 \varpi ^2 A K+5 \varpi ^2 B \lambda r^2 \eta+4 \varpi ^2 B \eta+\varpi ^2 B \mu \eta\nonumber\\ &&\nonumber\\ && \hspace{0.5cm}+3 B \lambda r^2,\\
&&\hspace{-0.5cm}\eta_{11}=\frac{\varpi ^2 B \lambda r}{1-K}+\frac{B \lambda r \left(\varpi ^2 \eta+1\right)}{(1-\mu) \eta}+A\Bigg[\frac{2 \varpi ^4 \lambda r \log \left(-\frac{C r^2+\mu}{\mu-1}\right)}{\mu-1}-\frac{2 \varpi ^2 \lambda r \eta}{\lambda r^2+\mu}-\frac{2 \varpi ^2 \lambda r}{(1-\mu) \eta}+\frac{2 \varpi ^2 \lambda r \log \left(\varpi ^2 \eta+1\right)}{(1-\mu) \eta}\nonumber\\ &&\nonumber\\ && \hspace{0.5cm}-\frac{2 \varpi ^2 \lambda r \log \left(\eta\right)}{(1-\mu) \eta} -\frac{2 \varpi ^6 \lambda r \left(\lambda r^2+\mu\right)}{(1-\mu) (\mu-1) \eta \left(\varpi ^2 \eta+1\right)}+\frac{2 \varpi ^4 \lambda r}{(1-\mu) \left(\varpi ^2 \eta+1\right)}-\frac{4 \varpi ^4 \lambda r \log \left(\varpi^2 \eta+1\right)}{\mu-1}+\frac{2 \varpi ^4 \lambda r}{\mu-1}\Bigg],\\
\end{eqnarray*}
\begin{eqnarray*}
&&\hspace{-0.5cm}\eta_{12}=(\mu-1)^2 \mu \left(\lambda r^2+1\right)^2 \eta^{3} \Bigg[A \Big(\frac{\varpi ^4 \left(\lambda r^2+\mu\right) \log \left(-\frac{\lambda r^2+\mu}{\mu-1}\right)}{\mu-1}-2 \varpi ^2 \eta-2 \alpha ^2 \eta \log \left(\eta\right)+2 \varpi ^2 \eta \log \left(\varpi ^2 \eta+1\right)\nonumber\\ &&\nonumber\\ && \hspace{0.5cm}-\frac{2 \varpi ^4 \left(\lambda r^2+\mu\right) \log \left(\varpi ^2 \eta+1\right)}{\mu-1}-1\Big)+B \eta \left(\varpi ^2 \eta+1\right)\Bigg]^2,\\
&&\hspace{-0.5cm}\eta_{13}=
\zeta_1 \lambda \left(\lambda r^2+\mu\right)^2 \Bigg[-\frac{2 \varpi ^2 A \lambda r \left(\lambda r^2 \left(5 \varpi ^2 \eta+3\right)+\varpi ^2 \mu \eta+4 \varpi ^2 \eta+\mu+2\right)}{\lambda r^2+\mu}+\nonumber\\ &&\nonumber\\ && \hspace{0.5cm}\frac{2 \varpi ^4 A \lambda r \left(\lambda r^2 \left(5 \varpi^2 \eta+3\right)+\varpi ^2 \mu \eta+4 \varpi ^2 \eta+\mu+2\right)}{(1-\mu) \eta \left(\varpi^2 \eta+1\right)}-\frac{A \lambda r}{(1-\mu) \eta}+\frac{A \lambda \mu r}{(1-\mu) \eta}-20 \varpi^2 A \lambda r+\frac{5 \varpi ^2 B \lambda^2 r^3}{(1-\mu) \eta}\nonumber\\ &&\nonumber\\ && \hspace{0.5cm}+10 \varpi ^2 B \lambda r \eta+\frac{4 \varpi ^2 B \lambda r}{(1-\mu) \eta}+\frac{\varpi ^2 B \lambda \mu r}{(1-\mu) \eta}+6 B \lambda r\Bigg],\\
&&\hspace{-0.5cm}\eta_{14}=
\varpi ^2 A \Bigg(\frac{5 \varpi ^2 \lambda^2 r^3}{(1-\mu) \eta}+2 \lambda r \left(5 \varpi ^2 \eta+3\right)+\frac{4 \varpi ^2 \lambda r}{(1-\mu) \eta}+\frac{\varpi ^2 \lambda \mu r}{(1-\mu) \eta}\Bigg) \log \left(-\frac{\lambda r^2+\mu}{\mu-1}\right)\nonumber\\ &&\nonumber\\ && \hspace{0.5cm}+2 \varpi ^2 A \Bigg(\frac{5 \varpi ^2 \lambda^2 r^3}{(1-\mu) \eta}+2 \lambda r \left(5 \varpi ^2 \eta+3\right)+\frac{4 \varpi ^2 \lambda r}{(1-\mu) \eta}+\frac{\varpi ^2 \lambda \mu r}{(1-\mu) \eta}\Bigg) \log \left(\varpi ^2 \eta+1\right)\Bigg),\\
&&\hspace{-0.5cm}\eta_{15}=(\mu-1)^2 \mu \left(\lambda r^2+1\right)^2 \left(\frac{\lambda r^2+\mu}{1-\mu}\right)^{3/2} \Bigg[A \Bigg(\frac{\varpi ^4 \left(\lambda r^2+\mu\right) \log \left(-\frac{\lambda r^2+\mu}{\mu-1}\right)}{\mu-1}-2 \varpi^2 \eta\nonumber\\ &&\nonumber\\ && \hspace{0.5cm}-2 \varpi ^2 \eta\log \left(\eta\right)+2 \varpi ^2 \eta\log \left(\varpi ^2 \eta+1\right)-\frac{2 \varpi ^4 \left(\lambda r^2+\mu\right) \log \left(\varpi^2 \eta+1\right)}{\mu-1}-1\Bigg)\nonumber\\ &&\nonumber\\ && \hspace{5.5cm}+B \eta\left(\varpi ^2 \eta+1\right)\Bigg],\\ 
&&\hspace{-0.5cm}\eta_{16}=3 \zeta_1 \lambda^2 r \left(\lambda r^2+\mu\right)^2 \Bigg[-8 \varpi ^2 A-\varpi ^2 A \left(\lambda r^2 \left(5 \varpi ^2 \eta+3\right)+\varpi ^2 \mu \eta+4 \varpi^2 \eta+\mu+2\right) \log \left(-\frac{\lambda r^2+\mu}{\mu-1}\right)\nonumber\\ &&\nonumber\\ && \hspace{1.5cm}+2 \varpi^2 A \left(\lambda r^2 \left(5 \varpi ^2 \eta+3\right)+\varpi ^2 \mu \eta+4 \varpi^2 \eta+K+2\right) \log \left(\varpi ^2 \eta+1\right)-A \eta+A \mu \eta-10 \varpi ^2 A \lambda r^2\nonumber\\ &&\nonumber\\ && \hspace{5.5cm} -2 \varpi ^2 A \mu+5 \varpi ^2 B \lambda r^2 \eta+4 \varpi ^2 B \eta+\varpi ^2 B \mu \eta+3 B \lambda r^2+B \mu+2 B\Bigg],
\end{eqnarray*}
\begin{eqnarray*}
&&\hspace{-0.5cm}\eta_{17}=(1-\mu) (\mu-1)^2 K \left(\lambda r^2+1\right)^2 \left(\frac{\lambda r^2+\mu}{1-\mu}\right)^{5/2} \Bigg[\lambda \Big(\frac{\varpi ^4 \left(\lambda r^2+\mu\right) \log \left(-\frac{\lambda r^2+\mu}{\mu-1}\right)}{\mu-1}-2 \varpi ^2 \eta-2 \varpi ^2 \eta \log \left(\eta\right)\nonumber\\ &&\nonumber\\ && \hspace{3.5cm}+2 \varpi ^2 \eta \log \left(\varpi ^2 \eta+1\right)-\frac{2 \varpi ^4 \left(\lambda r^2+\mu\right) \log \left(\varpi ^2 \eta+1\right)}{\mu-1}-1\Big)+\lambda \eta \left(\varpi ^2 \eta+1\right)\Bigg],\\
\end{eqnarray*}
\begin{eqnarray*}
&&\hspace{-0.5cm}\eta_{18}=4 \zeta_1 \lambda^2 r \left(\lambda r^2+\mu\right) \Bigg[-8 \varpi ^2 A-\varpi ^2 A \left(\lambda r^2 \left(5 \varpi ^2 \eta+3\right)+\varpi ^2 \mu \eta+4 \varpi ^2 \eta+\mu+2\right) \log \left(-\frac{\lambda r^2+\mu}{\mu-1}\right)\nonumber\\ &&\nonumber\\ && \hspace{0.5cm}+2 \varpi ^2 A \left(\lambda r^2 \left(5 \varpi ^2 \eta+3\right)+\varpi ^2 \mu \eta+4 \varpi^2 \eta+\mu+2\right) \log \left(\varpi ^2 \eta+1\right)-A \eta+A \mu \eta-10 \varpi ^2 A \lambda r^2\nonumber\\ &&\nonumber\\ && \hspace{5.5cm}-2 \varpi ^2 A \mu+5 \varpi ^2 B \lambda r^2 \eta+4 \varpi ^2 B \eta+\varpi ^2 B \mu \eta+3 B \lambda r^2+B \mu+2 B\Bigg],\\
&&\hspace{-0.5cm}\eta_{19}=(\mu-1)^2 K \left(\lambda r^2+1\right)^2 \left(\frac{\lambda r^2+\mu}{1-\mu}\right)^{3/2} \Bigg[A \Big(\frac{\varpi ^4 \left(\lambda r^2+\mu\right) \log \left(-\frac{\lambda r^2+\mu}{\mu-1}\right)}{\mu-1}-2 \varpi ^2 \eta\nonumber\\ &&\nonumber\\ && \hspace{2cm}-2 \varpi ^2 \eta \log \left(\eta\right)+2 \varpi^2 \eta \log \left(\varpi ^2 \eta+1\right)-\frac{2 \varpi ^4 \left(\lambda r^2+\mu\right) \log \left(\varpi ^2 \eta+1\right)}{\mu-1}-1\Big)+B \eta \left(\varpi ^2 \eta+1\right)\Bigg],\\
&&\hspace{-0.5cm}\eta_{20}=4 \zeta_1 \lambda^2 r \left(\lambda r^2+\mu\right)^2 \Bigg[-8 \varpi ^2 A-\varpi ^2 A \left(\lambda r^2 \left(5 \varpi^2 \eta+3\right)+\varpi^2 \mu \eta+4 \varpi ^2 \eta+\mu+2\right) \log \left(-\frac{\lambda r^2+\mu}{\mu-1}\right)\nonumber\\ &&\nonumber\\ && \hspace{2.5cm}+2 \varpi^2 A \left(\lambda r^2 \left(5 \varpi^2 \eta+3\right)+\varpi ^2 \mu \eta+4 \varpi ^2 \eta+\mu+2\right) \log \left(\varpi ^2 \eta+1\right)-A \eta+A \mu \eta\nonumber\\ &&\nonumber\\ && \hspace{4cm}-10 \varpi ^2 A \lambda r^2-2 \varpi^2 A \mu+5 \varpi ^2 B \lambda r^2 \eta+4 \varpi ^2 B \eta+\varpi ^2 B \mu \eta+3 B \lambda r^2+B \mu+2 B\Bigg],\\
&&\hspace{-0.5cm}\eta_{21}=(\mu-1)^2 \mu \left(\lambda r^2+1\right)^3 \left(\frac{\lambda r^2+\mu}{1-\mu}\right)^{3/2} \Bigg[A \Big(\frac{\varpi ^4 \left(\lambda r^2+\mu\right) \log \left(-\frac{\lambda r^2+\mu}{\mu-1}\right)}{\mu-1}-2 \varpi^2 \eta-2 \varpi ^2 \eta \log \left(\eta\right)\nonumber\\ &&\nonumber\\ && \hspace{4.5cm}+2 \varpi ^2 \eta \log \left(\varpi ^2 \eta+1\right)-\frac{2 \varpi ^4 \left(\lambda r^2+\mu\right) \log \left(\varpi^2 \eta+1\right)}{\mu-1}-1\Big)+B \eta \left(\varpi ^2 \eta+1\right)\Bigg],\\
&&\hspace{-0.5cm}\eta_{22}=-\frac{\zeta_1 \lambda^2 r^2 \Bigg[\frac{4 \varpi ^2 \lambda r}{(1-\mu) \left(\eta-\frac{\varpi ^2 \left(\lambda r^2+\mu\right)}{\mu-1}\right)}-\frac{2 \varpi ^2 \left(\lambda r^2+1\right) \left(\frac{\lambda r}{(1-\mu) \eta}-\frac{2 \varpi^2 \lambda r}{\mu-1}\right)}{(1-\mu) \left(\eta-\frac{\varpi^2 \left(\lambda r^2+\mu\right)}{\mu-1}\right)^2}-\frac{8 \lambda r \left(-3 \lambda r^2-5 \mu+2\right)}{\left(4 \lambda r^2+4\right)^2}-\frac{6 \lambda r}{4 \lambda r^2+4}\Bigg]}{2 \mu \left(\lambda r^2+1\right)^2}\nonumber\\ &&\nonumber\\ && \hspace{8.5cm}-\frac{8 \zeta_1 \lambda^2 (\mu-1) r}{\mu \left(\lambda r^2+1\right)^3}-\frac{2 \zeta_1 \lambda^2 (\mu-1) K r}{\left(\lambda \mu r^2+\mu\right)^2},\\
 &&\hspace{-1.2cm}\eta_{23}=\frac{2 \zeta_1 \lambda^3 r^3}{\mu \left(\lambda r^2+1\right)^3} \Bigg[\frac{2 \varpi ^2 \left(\lambda r^2+1\right)}{(1-\mu) \left(\eta-\frac{\varpi^2 \left(\lambda r^2+\mu\right)}{\mu-1}\right)}+\frac{-3 \lambda r^2-5 \mu+2}{4 \lambda r^2+4}+\mu-1\Bigg]\nonumber\\ &&\nonumber\\ && \hspace{6.2cm}-\frac{\zeta_1 \lambda^2 r \Bigg[\frac{2 \varpi^2 \left(\lambda r^2+1\right)}{(1-\mu) \left(\eta-\frac{\varpi ^2 \left(\lambda r^2+\mu\right)}{\mu-1}\right)}+\frac{-3 \lambda r^2-5 \mu+2}{4 \lambda r^2+4}+\mu-1\Bigg]}{\mu \left(\lambda r^2+1\right)^2},
 \end{eqnarray*}
\begin{eqnarray*}
&&\hspace{-0.5cm}\eta_{24}=\frac{\eta_{13}-\eta_{14}}{\eta_{15}-\eta_{16}},~~~~\eta_{25}=\frac{\eta_{19}-\eta_{20}}{(\eta_{17}-\eta_{18})\eta_{21}},~~~\Psi=\log \left(\varpi ^2 \sqrt{\frac{\lambda r^2+\mu}{1-\mu}}+1\right).\\
   &&\hspace{-0.5cm}\eta_{27}(R)= -8 \zeta_1 \lambda \mu+8 \zeta_1 \lambda \mu^2-8 \zeta_1 \lambda^2 R^2-14 \zeta_1 \lambda^2 \mu R^2+31 \zeta_1 \lambda^2 \mu^2 R^2-9 \zeta_1 \lambda^2 \mu^3 R^2-22 \zeta_1 \lambda^3 R^4+22 \zeta_1 \lambda^3 \mu R^4\\ &&\nonumber\\ && \hspace{0.4cm}+4 \zeta_1 \lambda^3 \mu^2 R^4-4 \zeta_1 \lambda^3 \mu^3 R^4-9 \zeta_1 \lambda^4 R^6+13 \zeta_1 \lambda^4 \mu R^6-4 \zeta_1 \lambda^4 \mu^2 R^6-88 \varpi ^2 \zeta_1 \lambda \mu \eta+88 \varpi ^2 \zeta_1 \lambda \mu^2 \eta\\ &&\nonumber\\ && \hspace{0.4cm}-96 \varpi ^2 \zeta_1 \lambda^2 R^2 \eta-98 \varpi ^2 \zeta_1 \lambda^2 \mu R^2 \eta+221 \varpi ^2 \zeta_1 \lambda^2 \mu^2 R^2 \eta-27 \varpi ^2 \zeta_1 \lambda^2 \mu^3 R^2 \eta-210 \varpi ^2 \zeta_1 \lambda^3 R^4 \eta\\ &&\nonumber\\ && \hspace{0.4cm}+146 \varpi ^2 \zeta_1 \lambda^3 \mu R^4 \eta+76 \varpi ^2 \zeta_1 \lambda^3 \mu^2 R^4 \eta-12 \varpi ^2 \zeta_1 \lambda^3 \mu^3 R^4 \eta-99 \varpi ^2 \zeta_1 \lambda^4 R^6 \eta+111 \varpi ^2 \zeta_1 \lambda^4 \mu R^6 \eta\\ &&\nonumber\\ && \hspace{0.4cm}-12 \varpi ^2 \zeta_1 \lambda^4 \mu^2 R^6 \eta-80 \varpi ^4 \zeta_1 \lambda K^2-176 \varpi ^4 \zeta_1 \lambda^2 \mu R^2-172 \varpi ^4 \zeta_1 \lambda^2 \mu^2 R^2+18 \varpi ^4 \zeta_1 \lambda^2 \mu^3 R^2-96 \varpi ^4 \zeta_1 \lambda^3 R^4\\ &&\nonumber\\ && \hspace{0.4cm}-376 \varpi ^4 \zeta_1 \lambda^3 \mu R^4-46 \varpi ^4 \zeta_1 \lambda^3 \mu^2 R^4+8 \varpi ^4 \zeta_1 \lambda^3 \mu^3 R^4-204 \varpi ^4 \zeta_1 \lambda^4 R^6-162 \varpi ^4 \zeta_1 \lambda^4 \mu R^6+16 \varpi^4 \zeta_1 \lambda^4 \mu^2 R^6\\ &&\nonumber\\ && \hspace{0.4cm}-98 \varpi ^4 \zeta_1 \lambda^5 R^8+8 \varpi ^4 \zeta_1 \lambda^5 \mu R^8-4 \zeta_2 \mu^2+4 \zeta_2 \mu^3-4 \zeta_2 \lambda \mu R^2-8 \zeta_2 \lambda \mu^2 R^2+12 \zeta_2 \lambda \mu^3 R^2-12 \zeta_2 \lambda^2 K R^4+\\ &&\nonumber\\ && \hspace{0.4cm}12 \zeta_2 \lambda^2 \mu^3 R^4-12 \zeta_2 \lambda^3 \mu R^6+8 \zeta_2 \lambda^3 \mu^2 R^6+4 \zeta_2 \lambda^3 \mu^3 R^6-4 \zeta_2 \lambda^4 \mu R^8+4 \zeta_2 \lambda^4 \mu^2 R^8-12 \varpi ^2 \zeta_2 K^2 \eta\\ &&\nonumber\\ && \hspace{0.4cm}+12 \varpi ^2 \zeta_2 \mu^3 \eta-12 \varpi ^2 \zeta_2 \lambda \mu R^2 \eta-24 \varpi^2 \zeta_2 \lambda \mu^2 R^2 \eta+36 \varpi^2 \zeta_2 \lambda \mu^3 R^2 \eta-36 \varpi^2 \zeta_2 \lambda^2 \mu R^4 \eta\\ &&\nonumber\\ && \hspace{0.4cm}+36 \varpi^2 \zeta_2 \lambda^2 \mu^3 R^4 \eta-36 \varpi ^2 \zeta_2 \lambda^3 \mu R^6 \eta+24 \varpi ^2 \zeta_2 \lambda^3 \mu^2 R^6 \eta+12 \varpi ^2 \zeta_2 \lambda^3 \mu^3 R^6 \eta-12 \varpi ^2 \zeta_2 \lambda^4 \mu R^8 \eta\\ &&\nonumber\\ && \hspace{0.4cm}+12 \varpi ^2 \zeta_2 \lambda^4 K^2 R^8 \eta-8 \varpi ^4 \zeta_2 \mu^3-16 \varpi ^4 \zeta_2 \lambda \mu^2 R^2-24 \varpi ^4 \zeta_2 \lambda \mu^3 R^2-8 \varpi^4 \zeta_2 \lambda^2 \mu R^4-48 \varpi ^4 \zeta_2 \lambda^2 \mu^2 R^4\\ &&\nonumber\\ && \hspace{0.4cm}-24 \varpi ^4 \zeta_2 \lambda^2 \mu^3 R^4-24 \varpi ^4 \zeta_2 \lambda^3 K R^6-48 \varpi^4 \zeta_2 \lambda^3 \mu^2 R^6-8 \varpi ^4 \zeta_2 \lambda^3 K^3 R^6-24 \varpi ^4 \zeta_2 \lambda^4 \mu R^8-16 \varpi^4 \zeta_2 \lambda^4 \mu^2 R^8\\ &&\nonumber\\ && \hspace{0.2cm}-8 \varpi ^4 \zeta_2 \lambda^5 \mu R^{10}-\varpi ^2 \left(\lambda R^2+\mu\right) \log \left(\eta\right) \Big(\varpi ^2 \lambda^4 R^6 \left(\zeta_1 \left(49 \varpi^2 \eta-4 \mu \left(\varpi ^2 \eta+2\right)+74\right)+4 \zeta_2 \mu R^2 \left(\varpi ^2 \eta+2\right)\right)\\ &&\nonumber\\ && \hspace{0.2cm}+2 \lambda^3 R^4 \Big(\zeta_1 \left(78 \varpi ^2-2 \varpi ^2 \mu^2 \left(\varpi ^2 \eta+2\right)+51 \varpi ^4 \eta+4 \mu \left(6 \varpi ^2+4 \varpi ^4 \eta-3 \eta\right)+12 \eta\right)+2 \varpi ^2 \zeta_2 \mu (\mu+3) R^2\\ &&\nonumber\\ && \hspace{0.2cm} \left(\varpi ^2 \eta+2\right)\Big )+\lambda^2 R^2 \Big(\zeta_1 \Big(-\left(\mu^2 \left(18 \varpi ^2+9 \varpi^4 \eta+8 \eta\right)\right)+2 \mu \left(70 \varpi ^2+43 \varpi ^4 \eta-16 \eta\right)\\ &&\nonumber\\ && \hspace{0.2cm}+8 \left(9 \varpi ^2+6 \varpi^4 \eta+5 \eta\right)\Big)+12 \varpi ^2 \zeta_2 \mu (\mu+1) R^2 \left(\varpi ^2 \eta+2\right)\Big)+4 \varpi ^2 \zeta_2 \mu^2 \left(\varpi^2 \eta+2\right)\\ &&\nonumber\\ && \hspace{0.2cm}+4 \lambda \Big(\varpi ^2 \zeta_2 K (3 \mu+1) R^2 \left(\varpi ^2 \eta+2\right)-2 \zeta_1 \left(K^2 \eta+\mu \left(-8 \varpi ^2-5 \varpi ^4 \eta+\eta\right)-2 \eta\right)\Big)\Big)\\ &&\nonumber\\ && \hspace{0.2cm}+2(1-\mu)\eta^{3}\varpi ^2\left(\lambda^3 R^4 \left(\zeta_1-4 \zeta_1 \mu+4 \zeta_2 \mu R^2\right)+\lambda^2 R^2 \left(14 \zeta_1-17 \zeta_1 \mu+12 \zeta_2 \mu R^2\right)+4 \zeta_2 \mu\right),  
\end{eqnarray*}

\begin{eqnarray*}
 &&\hspace{-0.5cm} \eta_{28}=\lambda \left(8 \zeta_1-8 \zeta_1 \mu+12 \zeta_2 \mu R^2\right)\log \left(\eta\right)+48 \varpi^2 \zeta_1 \lambda \mu \eta \Psi-48 \varpi^2 \zeta_1 \lambda \mu^2 \eta \Psi+48 \varpi ^2 \zeta_1 \lambda^2 R^2 \eta \Psi\\ &&\nonumber\\ && \hspace{0.2cm}+60 \varpi^2 \zeta_1 \lambda^2 \mu R^2 \eta \Psi-126 \varpi^2 \zeta_1 \lambda^2 \mu^2 R^2 \eta \Psi+18 \varpi^2 \zeta_1 \lambda^2 \mu^3 R^2 \eta \Psi+108 \varpi^2 \zeta_1 \lambda^3 R^4 \eta \Psi-76 \varpi^2 \zeta_1 \lambda^3 \mu R^4 \eta \Psi\\ &&\nonumber\\ && \hspace{0.2cm}-40 \varpi ^2 \zeta_1 \lambda^3 \mu^2 R^4 \eta \Psi+8 \varpi^2 \zeta_1 \lambda^3 \mu^3 R^4 \eta \Psi+50 \varpi^2 \zeta_1 \lambda^4 R^6 \eta \Psi-58 \varpi^2 \zeta_1 \lambda^4 \mu R^6 \eta \Psi+8 \varpi^2 \zeta_1 \lambda^4 \mu^2 R^6 \eta \Psi\\ &&\nonumber\\ && \hspace{0.2cm}+128 \varpi^4 \zeta_1 \lambda \mu^2 \Psi+272 \varpi^4 \zeta_1 \lambda^2 \mu R^2 \Psi+280 \varpi^4 \zeta_1 \lambda^2 \mu^2 R^2 \Psi-36 \varpi^4 \zeta_1 \lambda^2 \mu^3 R^2 \Psi+144\varpi^4 \zeta_1 \lambda^3 R^4 \Psi\\ &&\nonumber\\ && \hspace{0.2cm}+592 \varpi^4 \zeta_1 \lambda^3 K R^4 \Psi+60 \varpi^4 \zeta_1 \lambda^3 \mu^2 R^4 \Psi-16 \varpi^4 \zeta_1 \lambda^3 \mu^3 R^4 \Psi+312 \varpi ^4 \zeta_1 \lambda^4 R^6 \Psi+244 \varpi^4 \zeta_1 \lambda^4 \mu R^6 \Psi\\ &&\nonumber\\ && \hspace{0.2cm}-32\varpi ^4 \zeta_1 \lambda^4 \mu^2 R^6 \Psi+148 \varpi^4 \zeta_1 \lambda^5 R^8 \Psi-16 \varpi^4 \zeta_1 \lambda^5 \mu R^8 \Psi+80 \varpi ^6 \zeta_1 \lambda \mu^2 \eta \Psi+176 \varpi^6 \zeta_1 \lambda^2 \mu R^2 \eta \Psi\\ &&\nonumber\\ && \hspace{0.2cm}+172 \varpi^6 \zeta_1 \lambda^2 \mu^2 R^2 \eta \Psi-18 \varpi^6 \zeta_1 \lambda^2 \mu^3 R^2 \eta \Psi+96 \varpi^6 \zeta_1 \lambda^3 R^4 \eta \Psi+376 \varpi ^6 \zeta_1 \lambda^3 \mu R^4 \eta \Psi+46 \varpi ^6 \zeta_1 \lambda^3 \mu^2 R^4 \eta \Psi\\ &&\nonumber\\ && \hspace{0.2cm}-8 \varpi ^6 \zeta_1 \lambda^3 \mu^3 R^4 \eta \Psi+204 \varpi ^6 \zeta_1 \lambda^4 R^6 \eta \Psi+162 \varpi^6 \zeta_1 \lambda^4 \mu R^6 \eta \Psi-16 \varpi ^6 \zeta_1 \lambda^4 \mu^2 R^6 \eta \Psi+98 \varpi^6 \zeta_1 \lambda^5 R^8 \eta \Psi\\ &&\nonumber\\ && \hspace{0.2cm}-8 \varpi ^6 \zeta_1 \lambda^5 \mu R^8 \eta \Psi+8 \varpi ^2 \zeta_2 \mu^2 \eta \Psi-8 \varpi ^2 \zeta_2 \mu^3 \eta \Psi+8 \varpi ^2 \zeta_2 \lambda \mu R^2 \eta \Psi+16 \varpi^2 \zeta_2 \lambda \mu^2 R^2 \eta \Psi-24 \varpi ^2 \zeta_2 \lambda \mu^3 R^2 \eta \Psi\\ &&\nonumber\\ && \hspace{0.2cm}+24 \varpi ^2 \zeta_2 \lambda^2 \mu R^4 \eta \Psi-24 \varpi^2 \zeta_2 \lambda^2 \mu^3 R^4 \eta \Psi+24 \varpi ^2 \zeta_2 \lambda^3 \mu R^6 \eta \Psi-16 \varpi ^2 \zeta_2 \lambda^3 \mu^2 R^6 \eta \Psi-8 \varpi^2 \zeta_2 \lambda^3 \mu^3 R^6 \eta \Psi\\ &&\nonumber\\ && \hspace{0.2cm}+8 \varpi ^2 \zeta_2 \lambda^4 \mu R^8 \eta \Psi-8 \varpi ^2 \zeta_2 \lambda^4 \mu^2 R^8 \eta \Psi+16 \varpi ^4 \zeta_2 \mu^3 \Psi+32 \varpi ^4 \zeta_2 \lambda \mu^2 R^2 \Psi+48 \varpi ^4 \zeta_2 \lambda \mu^3 R^2 \Psi\\ &&\nonumber\\ && \hspace{0.2cm}+16 \varpi ^4 \zeta_2 \lambda^2 \mu R^4 \Psi+96 \varpi^4 \zeta_2 \lambda^2 \mu^2 R^4 \Psi+48 \varpi^4 \zeta_2 \lambda^2 \mu^3 R^4 \Psi+48 \varpi^4 \zeta_2 \lambda^3 \mu R^6 \Psi+96 \varpi ^4 \zeta_2 \lambda^3 \mu^2 R^6 \Psi\\ &&\nonumber\\ && \hspace{0.2cm}+16 \varpi^4 \zeta_2 \lambda^3 \mu^3 R^6 \Psi+48 \varpi ^4 \zeta_2 \lambda^4 \mu R^8 \Psi+32 \varpi^4 \zeta_2 \lambda^4 \mu^2 R^8 \Psi+16 \varpi ^4 \zeta_2 \lambda^5 \mu R^{10} \Psi+8 \varpi^6 \zeta_2 \mu^3 \eta \Psi\\ &&\nonumber\\ && \hspace{0.2cm}+16 \varpi ^6 \zeta_2 \lambda \mu^2 R^2 \eta \Psi+24 \varpi ^6 \zeta_2 \lambda \mu^3 R^2 \eta \Psi+8 \varpi^6 \zeta_2 \lambda^2 \mu R^4 \eta \Psi+48 \varpi ^6 \zeta_2 \lambda^2 \mu^2 R^4 \eta \Psi+24 \varpi ^6 \zeta_2 \lambda^2 \mu^3 R^4 \eta \Psi\\ &&\nonumber\\ && \hspace{0.2cm}+24 \varpi ^6 \zeta_2 \lambda^3 \mu R^6 \eta \Psi+48 \varpi ^6 \zeta_2 \lambda^3 \mu^2 R^6 \eta \Psi+8 \varpi ^6 \zeta_2 \lambda^3 \mu^3 R^6 \eta \Psi+24 \varpi ^6 \zeta_2 \lambda^4 \mu R^8 \eta \Psi+16 \varpi ^6 \zeta_2 \lambda^4 \mu^2 R^8 \eta \Psi\\ &&\nonumber\\ && \hspace{0.2cm}+8 \varpi ^6 \zeta_2 \lambda^5 \mu R^{10} \eta \Psi,\\
 &&\hspace{-0.5cm}\eta_{26}=  \lambda^4 \varpi ^2 \left(4 \mu \left(\eta \varpi ^2+2\right) \zeta_2 R^2+\left(49 \eta \varpi ^2-4 \mu \left(\eta \varpi ^2+2\right)+74\right) \zeta_1\right) R^6+4 \mu \left(\mu \left(\eta \varpi ^4+2 \varpi ^2-\eta\right)+\eta\right) \zeta_2\nonumber\\ &&\nonumber\\ && \hspace{0.5cm}+4 \lambda \left(\mu \left(3 \mu \eta \varpi ^4+\eta \varpi ^4+6 \mu \varpi ^2+2 \varpi ^2-3 \mu \eta+3 \eta\right) \zeta_2 R^2+2 \left(5 \mu \eta \varpi ^4+8 \mu \varpi^2-3 \mu \eta+3 \eta\right) \zeta_1\right)\nonumber\\ &&\nonumber\\ && \hspace{0.5cm}+\lambda^2 \Big(12 \mu \left(\eta \varpi ^4+2 \varpi^2+\mu \left(\eta \varpi ^4+2 \varpi ^2-\eta\right)+\eta\right) \zeta_2 R^4+\Big(-9 \left(\eta \varpi ^4+2 \varpi ^2-\eta\right) \mu^2+\nonumber\\ &&\nonumber\\ && \hspace{0.1cm}\left(86 \eta \varpi ^4+140 \varpi ^2-63 \eta\right) \mu+6 \left(8 \eta \varpi ^4+12 \varpi ^2+9 \eta\right)\Big) \zeta_1 R^2\Big)+\lambda^3 \Big(4 \mu \left(3 \eta \varpi ^4+6 \varpi ^2+\mu \left(\eta \varpi^4+2 \varpi ^2-\eta\right)\right) \nonumber\\ &&\nonumber\\ && \hspace{1.2cm} \zeta_2 R^6+\left(102 \eta \varpi ^4+156 \varpi ^2-4 \mu^2 \left(\eta \varpi ^4+2 \varpi ^2-\eta\right)+\mu \left(32 \eta \varpi ^4+48 \varpi ^2-29 \eta\right)+25 \eta\right) \zeta_1 R^4\Big),
 \end{eqnarray*}
\end{small}

\acknowledgments

S. K. Maurya appreciates the administration of the University of Nizwa in the Sultanate of Oman for their unwavering support and encouragement. Sourav Chaudhary expresses his gratitude to the Central University of Haryana for providing a University Research Fellowship~(URF) under the Reg. No. 222019. Jitendra Kumar is highly thankful to the Department of Mathematics, Central University of Haryana. Sweeti Kiroriwal acknowledges the University Grant Commission~(UGC), New Delhi, India under the NTA Ref. No. 211610000030 for providing financial support.

\paragraph{Data Availability Statement}
All required data is available with this manuscript.\\
\paragraph{Conflict of Interest Statement}
The authors declare that they have no conflict of interest personal relationships that could have appeared to influence the work reported in this paper.








\end{document}